\documentclass[smallextended]{svjour3}       % onecolumn (second format)
\usepackage{graphicx}
\usepackage{xspace}
\usepackage{xcolor}
\usepackage[authoryear]{natbib}
\usepackage{amsmath}
\usepackage{amssymb}
\usepackage{multirow}
\usepackage[leftcaption]{sidecap}
\usepackage{rotating}
\usepackage{lineno}
\usepackage{etoolbox}
\newcommand*{\affaddr}[1]{#1} % No op here. Customize it for different styles.
\newcommand*{\affmark}[1][*]{\textsuperscript{#1}}

\newcommand{\Fe}{{$^{55}$Fe\ }}
\newcommand{\Po}{{$^{210}$Po\ }}
\newcommand{\Ka}{{K$_{\alpha}$\ }}
\newcommand{\Kb}{{K$_{\beta}$\ }}
\newcommand{\OSIRIS}{{\it OSIRIS-REx}\xspace}
\newcommand{\NICER}{{\it NICER}\xspace}
\newcommand{\MAXI}{{\it MAXI}\xspace}
\newcommand{\NuSTAR}{{\it NuSTAR}\xspace}

\begin{document}

\title{Calibration and Performance of the REgolith X-Ray Imaging Spectrometer (REXIS) Aboard NASA's OSIRIS-REx Mission to Bennu}
%\title{Insert your title here%\thanks{Grants or other notes
%placed here. General acknowledgments should be placed at the end of the article.}}

\titlerunning{Calibration and Performance of REXIS on board \OSIRIS}

\author{Jaesub Hong\protect\affmark[1] \and
Richard P. Binzel\affmark[2] \and
Branden Allen\affmark[1] \and
David Guevel\affmark[1] \and
Jonathan Grindlay\affmark[1] \and
Daniel Hoak\affmark[1] \and
Rebecca Masterson\affmark[2] \and
Mark Chodas\affmark[2] \and
Madeline Lambert\affmark[2] \and
Carolyn Thayer\affmark[2] \and
Ed Bokhour\affmark[2]  \and
Pronoy Biswas\affmark[2]  \and
Jeffrey A. Mendenhall\affmark[3] \and 
Kevin Ryu\affmark[3] \and
James Kelly\affmark[3] \and
Keith Warner\affmark[3] \and
Lucy F. Lim\affmark[4] \and
Arlin Bartels\affmark[4]   \and
Dante S. Lauretta\affmark[5]  \and
William V. Boynton\affmark[5] \and
Heather L. Enos\affmark[5] \and
Karl Harshman\affmark[5] \and
Sara S. Balram-Knutson\affmark[5] \and
Anjani T. Polit\affmark[5] \and
Timothy J. McCoy\affmark[6] \and 
Benton C. Clark\affmark[7] 
}

\authorrunning{J.~Hong et~al.}

\institute{\email{jhong@cfa.harvard.edu} \\ 
    \affaddr{\affmark[1]Harvard University, Harvard-Smithsonian Center for Astrophysics, 60 Garden St, Cambridge, MA, USA, 02138} \\
    \affaddr{\affmark[2]Massachusetts Institute of Technology, 77 Massachusetts Ave, Cambridge, MA, USA} \\ \affaddr{\affmark[3]Massachusetts Institute of Technology Lincoln Laboratory, Lexington, MA, USA} 
    %\FIX[Requires LL Release Review prior to becoming a public document (Review in Progress)]\\
    \affaddr{\affmark[4]NASA Goddard Space Flight Center, Greenbelt, MD, USA} \\ \affaddr{\affmark[5] University of Arizona, Tucson, AZ, USA} \\ \affaddr{\affmark[6]Smithsonian Institution National Museum of Natural History, Washington, DC, USA.} \\ \affaddr{\affmark[7]Space Science Institute, Boulder, CO USA}
}

\date{Received: date / Accepted: date}

\maketitle

\begin{abstract}
The REgolith X-ray Imaging Spectrometer (REXIS) instrument on board NASA's \OSIRIS mission to the asteroid Bennu is a Class-D student collaboration experiment designed to detect fluoresced X-rays from the asteroid's surface to measure elemental abundances.  In July and November 2019 REXIS  collected $\sim$615 hours of integrated exposure time of Bennu's sun-illuminated surface from terminator orbits.  As reported in \citet{Hoak21}, the REXIS data do not contain a clear signal of X-ray fluorescence from the asteroid, in part due to the low incident solar X-ray flux during periods of observation.  To support the evaluation of the upper limits on the detectable X-ray signal that may provide insights for the properties of Bennu's regolith, we present an overview of the REXIS instrument, its operation, and details of its in-flight calibration on astrophysical X-ray sources.  This calibration includes the serendipitous detection of the transient X-ray binary MAXI J0637-430 during Bennu observations, demonstrating the operational success of REXIS at the asteroid.  We convey some lessons learned for future X-ray spectroscopy imaging investigations of asteroid surfaces.
\keywords{asteroids \and x-ray fluorescence \and x-ray astrophysics}
\end{abstract}

\section{Introduction}\label{sec:introduction}
In this paper, we report on the calibration and performance of the Regolith X-ray Imaging Spectrometer (REXIS) instrument aboard NASA's \OSIRIS asteroid sample return mission, which launched in 2016 and arrived at the asteroid Bennu in 2018.  As detailed in \citet{Masterson18}, REXIS was competitively selected as a Class-D student collaboration experiment to complement the remote sensing payload of \OSIRIS.  Class-D instruments are defined as having medium- to-low complexity, short mission operation lifetimes, and relatively low cost, while  providing platforms for technological innovation and training grounds for a diverse set of scientists and engineers. As reported by \citet{Hoak21}, the 2019 REXIS measurements %%at Bennu 
do not contain a clear signal of fluoresced X-rays from the surface of Bennu.  
%% fix this sentence:
%%Thus full analysis of the data, including possible implications for the asteroid surface characteristics that may come from the limits of detectability, 
Establishing upper limits on the fluxes of fluoresced X-rays
requires a careful description and analysis of the instrument calibration and in-flight performance. These upper limits may shed light on the surface characteristics of Bennu and help inform the design of future planetary X-ray experiments.

%%For
In this paper on REXIS calibration and in-flight performance,  we first give a brief overview of the REXIS instrument, its on-board signal processing, and pre-flight milestones and activities that established the instrument status at launch (Section \ref{sec:instrument}).  We follow with details of in-flight testing of the main X-ray Imaging Spectrometer (XIS), including the monitoring of its in-flight condition (Section \ref{sec:calibration}).  Well-known astrophysical targets were observed as part of the in-flight calibration of the imaging performance (Section \ref{sec:imaging}). During the period when REXIS observed the asteroid Bennu, the serendipitous detection of a transient X-ray binary provided synchronous confirmation of the instrument's sensitivity for detecting fluoresced X-rays from Bennu's regolith (Section \ref{s:maxi}).  We describe the in-flight calibration and characteristics of the Solar X-ray Monitor (SXM)  in Section \ref{sec:sxm}. 
Finally, we discuss lessons that may be applied to future X-ray flight experiments (Section \ref{sec:summary}).

\section{The REXIS Instrument}\label{sec:instrument}

\begin{SCfigure}
    \small
	\caption{X-ray Imaging Spectrometer (XIS) of REXIS is a coded-aperture telescope with a stainless mask and a 2 $\times$ 2 array of X-ray CCDs. The inset shows
	the Solar X-ray Monitor (SXM) of REXIS equipped with an SDD in a collimator and the readout electronics.}		
	\includegraphics[width=3.5in]{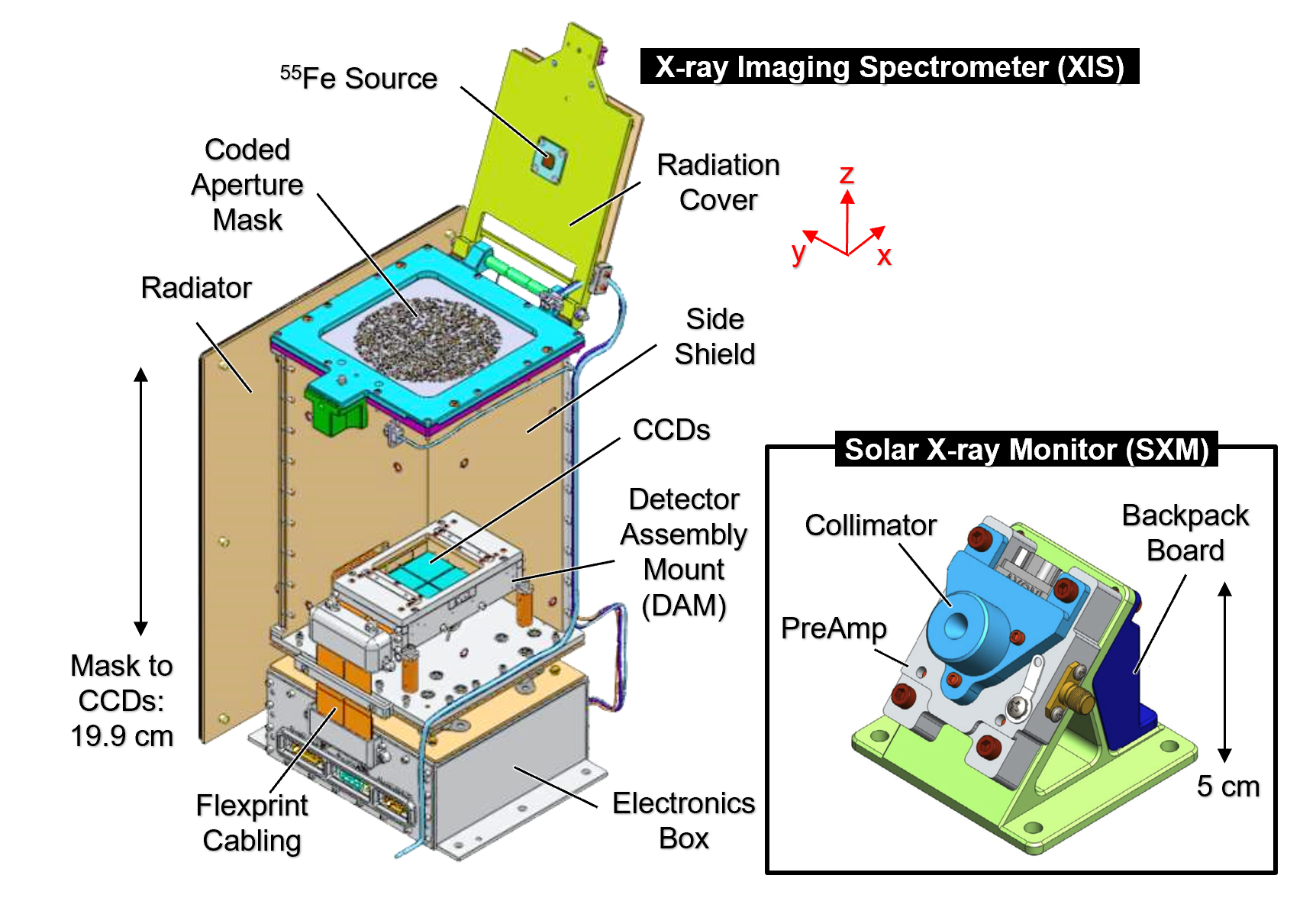}
	\label{f:instrument}
\end{SCfigure}

REXIS is an X-ray spectrometer designed to take advantage of the incident solar X-ray flux that can generate a diagnostic fluorescence signature from an asteroid’s surface. REXIS development built upon the experience of previously flown X-ray fluorescence (XRF) experiments dating back to the Apollo era and previous asteroid missions \citep[e.g.,][]{2000Sci...289.2101T, Nittler01, Okada06}.  We refer the reader to \citet{Masterson18} for a detailed description of the REXIS instrument.  
%%Here we summarize briefly:  
In brief summary here, REXIS consists of two components: a main X-ray Imaging Spectrometer (XIS) to measure XRF from Bennu and a separate Solar X-ray Monitor (SXM) to measure the incident X-ray flux from the Sun.  Figure~\ref{f:instrument} illustrates the key components of the REXIS instrument and Table~\ref{t:intrument} summarizes the key instrument parameters.

\begin{SCtable}[0.5]
%\centering
\small
\caption{Key Instrument Parameters of REXIS, based on the in-flight calibrations. See also
the pre-flight design values in  Table 1 in \citet{Masterson18}. Values in italic indicate changes relative to the original design.}
\begin{tabular}{ll}
\hline\hline
\bf Parameters & \bf Values \\
\hline
\multicolumn{2}{l}{\bf X-ray Imaging Spectrometer (XIS)} \\
Total Mass \& Power     & 6.5 kg \& 12.4W \\
Mask to Detector$^a$    & {\it 19.1} cm   \\
Field of View           & 28$^\circ$ (FWHM), 53$^\circ$ (FWZI$^f$) \\
Angular Resolution$^{a,b}$  & 26.5$'$ (5.6 m @ 730 m) \\
Localization Error$^{a,b}$  & $\sim$ 1$'$ -- 5$'$ \\
\hline
\multicolumn{2}{l}{\bf Detector in XIS: 2 $\times$ 2 CCDs} \\
Pixel size              & 24 $\mu$m $\times$ 24 $\mu$m \\
Depletion Depth         & 45 $\mu$m \\
Optical Blocking Filter (OBF)                     & 320 nm Al \\
%Total Area             & 25 cm$^2$ from 16 nodes \\
Active Area$^c$         & {\it 7.5} cm$^2$  \\
Effective Area$^c$      & {\it 4.4} cm$^2$ @ 1.25 keV (Mg-K) \\
                        & {\it 1.9} cm$^2$ @ 2.30 keV (S-K) \\
Energy Resolution$^c$   & {\it $\sim$150--300}~eV FWHM @ 5.9 keV \\
Energy Range$^d$        & {\it 0.4--10} keV \\
\hline
\multicolumn{2}{l}{\bf Coded Mask in XIS: 9.8 cm dia.~Stainless Steel} \\
Pixel pitch             & 1.536 mm (100 $\mu$m support grid)\\
Thickness               & 100 $\mu$m \\
Throughput              & 41\% with 50\% open pixels\\
\hline
\multicolumn{2}{l}{\bf Solar X-ray Monitor (SXM)} \\
Detector                & Amptek XR-100 SDD\\
Effective Area          & {\it 0.6 mm$^2$} @ 2.0 keV \\
                        & {\it 0.8 mm$^2$} @ 5.9 keV \\
Field of View           & {\it 42$^\circ$} (FWMI$^e$) {\it 60$^\circ$} (FWZI$^f$)\\
Energy Range$^g$        & \it $\sim$1.5 -- 20 keV \\
Optical Blocking        & 0.5 mil Be window \\
\hline
\multicolumn{2}{l}{$^a$Based on the Crab Calibration Operation (Section~\ref{sec:boresight}).}\\
\multicolumn{2}{l}{$^b$See Section~\ref{s:ang_resolution}.}\\
\multicolumn{2}{l}{$^c$Based on 5 stable nodes: see also Table~\ref{t:resolution}.}\\
\multicolumn{2}{l}{$^d$Constrained by the low energy excess (Section~\ref{sec:excess}).}\\
\multicolumn{2}{l}{$^e$Full Width Maximum Intensity}\\
\multicolumn{2}{l}{$^f$Full Width Zero Intensity} \\
\multicolumn{2}{l}{$^g$Nominal range, see Section~\ref{sec:sxm}}
\end{tabular}
\label{t:intrument}
\end{SCtable}

The XIS employs a Detector Assembly Mount (DAM) consisting of 4 MIT Lincoln Laboratory CCID-41 charge-coupled devices (CCDs) in a 2~$\times$~2 array \citep{Bautz04}.  
Each CCD is further divided into 4 separate readout channels, which we refer to as “nodes”, with each node consisting of 256~$\times$~1024 pixels of dimension 24 microns.   The CCID-41s are sensitive to the energy range of 0.4 to 10~keV with a spectral resolution (FWHM) of $\lesssim$ 220 eV at 5.9 keV. The primary fluoresced lines REXIS is designed to detect are Fe-L at 0.71 keV,  Mg-K at 1.25 keV, Al-K at 1.49 keV, Si-K at 1.74 keV, and S-K at 2.31 keV. 

REXIS includes a coded-aperture mask \citep{Dicke68}, unique to planetary science X-ray detectors, to facilitate image reconstruction of relative variations in elemental abundance across the surface of the asteroid. The REXIS mask contains a random pattern of holes with a throughput of 41\%.  The REXIS main detector array was protected from background radiation during the cruise phase by a cover that was successfully opened in September 2018, after approximately two years into flight. 

REXIS' second component, the SXM, utilizes an Amptek XR-100 Silicon Drift Detector (SDD) with a collimated effective area of about 0.8~mm$^2$, providing spectral coverage over a nominal energy range of 1.5 to 20 keV. With a large field of view (FWZI: 60$^\circ$), the SXM design enables monitoring of solar X-rays during nadir pointing of the XIS near or at terminator orbits of Bennu.

The Main Electronics Board (MEB), located beneath the REXIS XIS, serves to power and control the XIS and SXM. It also processes the XIS and SXM science data, and collects the instrument housekeeping (HK) data for downlink.

\subsection{REXIS On-board Signal Processing}

%% awkward sentence - could use a rewrite...
%% Requirements on the limited resource of 
Because spacecraft resources are finite, REXIS operations were required to fit within specific allocations of on-board data storage and data download volume.  Thus for all flight operations and outcomes we are reporting, it was necessary to process the full CCD images (frames) on-board. \cite{Allen13} and \cite{Jones14} describe the implementation of the REXIS firmware, and details of the REXIS software are given by \citet{Biswas16}. In brief, during REXIS observations the on-board software identified X-ray photon detections (excess charge above a preset threshold) in individual pixels and cataloging those photon detections as CCD “events”, so that only lists of CCD events (“event lists”) are downlinked to Earth instead of full frame images. Along with the CCD event list, instrument housekeeping (HK) parameters and the SXM histograms (see below) are also downlinked to Earth. 

During nominal REXIS operations, the CCDs collected 4-second exposures. At the end of each exposure the charge from each pixel is quickly (within 10 ms) transferred to a non-imaging framestore area of the detector before readout by the detector electronics. The framestore region is read serially into a single large array by sixteen 12-bit analog to digital converters (ADCs), one for each CCD node. As the data are read out, the firmware executes four steps of image processing: frame processing (e.g.~subtracting the noise pattern using a bias map and eliminating known bad pixels), event grading, event filtering, and finally science packet formation. The final science packet that is downlinked to Earth includes an X-ray event list giving the pixel location,  energy, and grade of each putative X-ray event. See Figure 11 in \citet{Masterson18} for the flow chart of the REXIS Image Processing Pipeline.  

Of particular importance to REXIS calibration and in-flight performance is the “grading” of individual photon detections. When an X-ray interaction occurs near the edge of a pixel in a CCD, the induced charge cloud can be distributed among the neighboring pixels and not easily recognized as a part of the discrete event. X-ray events are {\it graded} based on the degree of the charge splitting using a pre-set ``split-charge  threshold" (ST). Neighboring pixels above the ST are considered a valid signal, and their charge is added back to the sum of the signal. Depending on the proximity of the X-ray interaction to the pixel edge, the depth of the interaction in the Si layer and the electron cloud size (i.e., the energy of the incident X-ray), the split charge may be divided among as many as four pixels. 

During the event grading, each candidate detection is assigned a grade from 0 to 7 to identify how the charge from a single incident X-ray photon was divided among multiple pixels.  A Grade 0 event corresponds to an event localized in a single pixel (with no neighboring pixels above the split-charge threshold), while Grades 1 and higher denote events where charge is shared among one or more pixels adjacent to the central pixel.  Multi-pixel events generally yield poorer spectral resolution relative to single pixel events, since the combined electric noise and charge transfer inefficiency from multiple pixels contribute to the final signal sum. 

After the event grading algorithm, the flight software produces a list of valid CCD events for downlink. An event filtering algorithm ensures that the science data volume would not exceed downlink allocations during unexpected occurrences of high event rates due to factors such as solar flares, optical light leaks, hot pixels, or software glitches. The event rate filter was designed to randomly select at most 1000 (an adjustable parameter) events for downlink over each 4-second readout cycle.

The REXIS software was capable of downlinking a full or partial frame image of the XIS detector plane for a 4 second exposure, or the bias map of the entire plane (noise map). Downlinking frame images was a slow process, taking about 30 minutes for an entire detector plane image or 2 minutes for a full node image. Therefore, REXIS only downlinked full or partial frame and bias images when they were necessary for diagnostic purposes or other detailed analyses.

Unlike the fixed readout cycle in the REXIS CCDs, the SDD in the solar X-ray Monitor (SXM) responded continuously to incoming X-rays. The SXM readout electronics triggered on signals above a preset threshold. Given the potential for high solar X-ray flux during flares, REXIS did not downlink the individual SXM events, but rather a histogram that recorded the number of events observed in given energy bands. The SXM histogram had 512 bins with an energy spacing of about 59~eV and a maximum energy of 24~keV, and accumulated events over an integration period that alternated between 18 and 32 seconds. The alternating integration period was due to a software bug, but it had the beneficial effect of providing robustness to periods of high solar X-ray activity.

In addition to the REXIS science data, the MEB collected various housekeeping (HK) information from the REXIS system, and downlinked each HK packet every 64 seconds. The HK data included the voltage and current measurements of the CCD input biases, X-ray event rates (even when the CCD event data is not being downlinked), event filtering status, the output of several temperature sensors, and other diagnostic information.

\subsection{REXIS Pre-flight Challenges}

\citet{Masterson18} described the pre-flight activities of the REXIS main spectrometer and solar X-ray monitor.  As is the case for any instrument, the pre-flight activities (and challenges) shaped the status of the flight hardware and its operation in space.

In late 2014 and early 2015, Lincoln Laboratory delivered 8 CCDs to be used for the flight model (FM) and flight spare Detector Assembly Mounts (DAMs). Based on the measured performance of each of the CCDs (susceptibility to light leak, dark current level, and surface cosmetics), 4 CCDs were ranked Grade A, 2 CCDs Grade B, and the remaining 2 CCDs Grade C, where Grade A was considered to be of the highest quality. To mitigate risk, the REXIS team distributed the the CCDs between the flight model and flight spare detector assemblies: we assigned two Grade A CCDs to each assembly, two Grade B CCDs to the FM DAM, and the remaining two Grade C CCDs to the flight spare DAM. 

Dividing the CCDs between the two assemblies proved to be prudent, as the original FM DAM suffered contamination during a bake-out operation in a thermal vacuum chamber. Consequently, the flight spare DAM (with the two Grade C CCDs) was ultimately used as the flight instrument. The two Grade A CCDs used in the flight instrument were labeled as CCDs 1 and 3, and the Grade C CCDs were CCDs 0 and 2 (see Section \ref{sec:calibration}).  A further complication arose during vibration testing of the newly promoted flight model DAM, when a tear in a flexprint cable disabled signal connection from CCD 2 Node 3 and all nodes from CCD 3. Thus, at launch, the REXIS XIS had ten functioning nodes out of the 16 nodes in the detector assembly.  It was structurally necessary to leave the torn flexprint cable in place (but sealed to prevent further tearing), and this contributed to elevated electronic noise in the odd-numbered nodes (see Section~\ref{s:response}). 

For the SXM, during pre-launch testing we discovered that the trigger circuit for the SDD signals in the main electronics board was not properly connected. In addition, the MEB was found to suffer from ground-loop noise. We resolved these issues by introducing an additional electronics module (the "backpack board") that also served to condition the power for the Thermo-electric cooler (TEC) for the SDD. 

Due to these issues and other schedule delays, many aspects of the REXIS pre-flight ground calibration program were not fully completed, for example measurements of the CCD detection efficiency and energy resolution with external X-ray sources. Luckily, it was possible to perform pre-flight performance checks and troubleshooting using self-calibration of X-ray events from the onboard \Fe sources during thermal testing. In the case of the SXM, we performed several laboratory X-ray measurements using an external \Fe source to validate its functionality prior to flight.

\begin{table}
%\vspace{7cm}
\small
\caption{REXIS in-flight calibrations and observational activities}
\centering
\begin{tabular}{ lll }
\hline\hline
Activity & Start Date:  Duration* & Objectives \\
& (yy-mm-dd) & \\
\hline
Launch                      & 16-09-08  \\
L+14 days                   & 16-09-21:   $<$1 day  & System check: 14 days after launch \\
L+6 months                  & 17-03-14:   $<$1 day  & System check: 6 months after launch \\
L+12 months                 & 17-08-03:   $<$1 day  & System check: 12 months after launch \\
pre-EGA calib.              & 17-08-13:   4 days    & SXM calibration prior to EGA \\
post-EGA calib.             & 17-10-09:   4 days    & SXM calibration following EGA \\
L+18 months                 & 18-03-13:   $<$1 day  & System check: 18 months after launch \\
\bf Internal calib.         & 18-07-12:   4 days    & Calibrate the CCDs; measure internal background \\
L+22 months                 & 18-07-20:   2 days    & System check: 22 months after launch \\
Cover opening               & 18-09-14:   $<$1 day  & Open the radiation cover \\
\bf CXB calib.              & 18-10-09:   2 day     & Measure Cosmic X-ray Background (CXB) \\
Crab calib.~1               & 18-11-14:   4 days    & Calibrate the boresight and CCD Q.E.  \\ 
L+30 months                 & 19-02-16:   2 days    & System check: 30 months after launch \\
\bf Crab calib.~2           & 19-03-16:   4 days    & Calibrate the boresight and CCD Q.E. \\
\bf OBF calib.              & 19-04-18:   2 days    & Measure pin-holes in the Optical Blocking Filter (OBF) \\
\bf Mask calib.             & 19-06-02:   $<$1 day  & Measure the mask distortion using Sco X-1 \\
Orbital B obs.              & 19-07-01:   36 days   & Measure X-rays from Bennu during Orbital B \\
Orbital R obs.              & 19-11-11:   15 days   & Measure X-rays from Bennu during Orbital R \\
\hline
\multicolumn{3}{l}{*The data taking period per day ranges from a few to 16 hours}
\end{tabular}
\label{t:activity}
\end{table}

\section{Inflight Calibration and Response of X-ray Imaging Spectrometer}\label{sec:calibration}

Starting shortly after launch, REXIS carried out a series of calibration operations that collected data from on-board \Fe sources (described in the next section) and from observations of bright celestial X-ray sources. In addition, REXIS and the rest of the instruments onboard \OSIRIS were powered on for periodic checkups roughly every six months during the journey to Bennu. These activities were designed to characterize and monitor the instrument response over time. 

Table~\ref{t:activity} lists the operational activities of REXIS following the launch of \OSIRIS in 2016. Notable calibration operations included the 4-day internal background and calibration (with the radiation cover closed), the two sets of 4-day observations of the Crab Nebula for boresight and quantum efficiency (Q.E.) calibrations, the 1-day Optical Blocking Filter (OBF) calibration, and the 1-day observation of Sco X-1 for mask calibration. Throughout flight, REXIS used the 5.9~keV Mn-\Ka and 6.5~keV Mn-\Kb X-ray lines from the onboard \Fe sources to monitor changes in the spectral gain, offset in energy calibration, and change in Q.E. of the CCDs. As noted above, REXIS had a radiation cover located at the top of XIS tower above the coded mask, to minimize the radiation damage in the X-ray CCDs from cosmic ray interactions during the cruise phase to Bennu. The radiation cover was opened in September 2018, 8 months prior to the arrival of \OSIRIS at Bennu.

\subsection{Monitoring REXIS X-ray CCD Response using Onboard \Fe Sources} \label{sec:Fesources}

\begin{SCfigure}
    	%\centering
    	\small
		\includegraphics[width=3.0in]{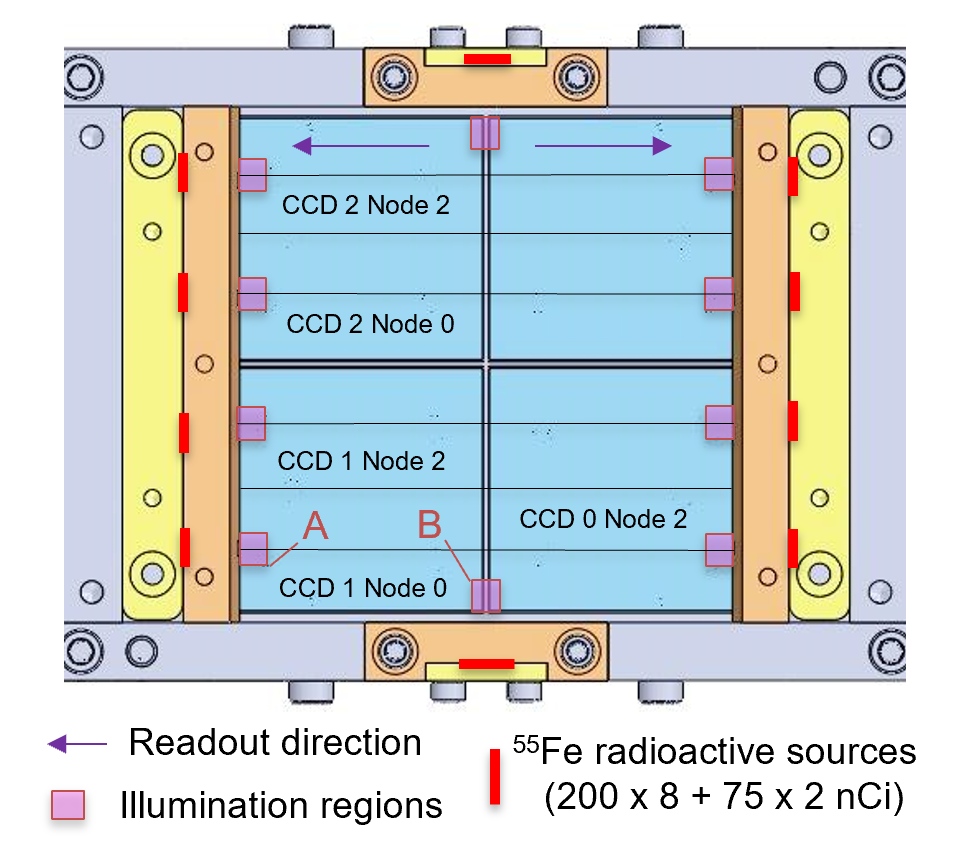}
		\caption{Onboard \Fe sources (red) and their illumination regions (purple) on the REXIS DAM. Five nodes with relatively stable performance are labeled. Labels A and B mark the near and far-end of the readout chain  exposed to the \Fe sources for CCD 1 node 0. }
			\label{f:Fe_sources}
\end{SCfigure}

\begin{SCfigure}
    \small
	\includegraphics[width=3.1in,trim=0 0 0 0]{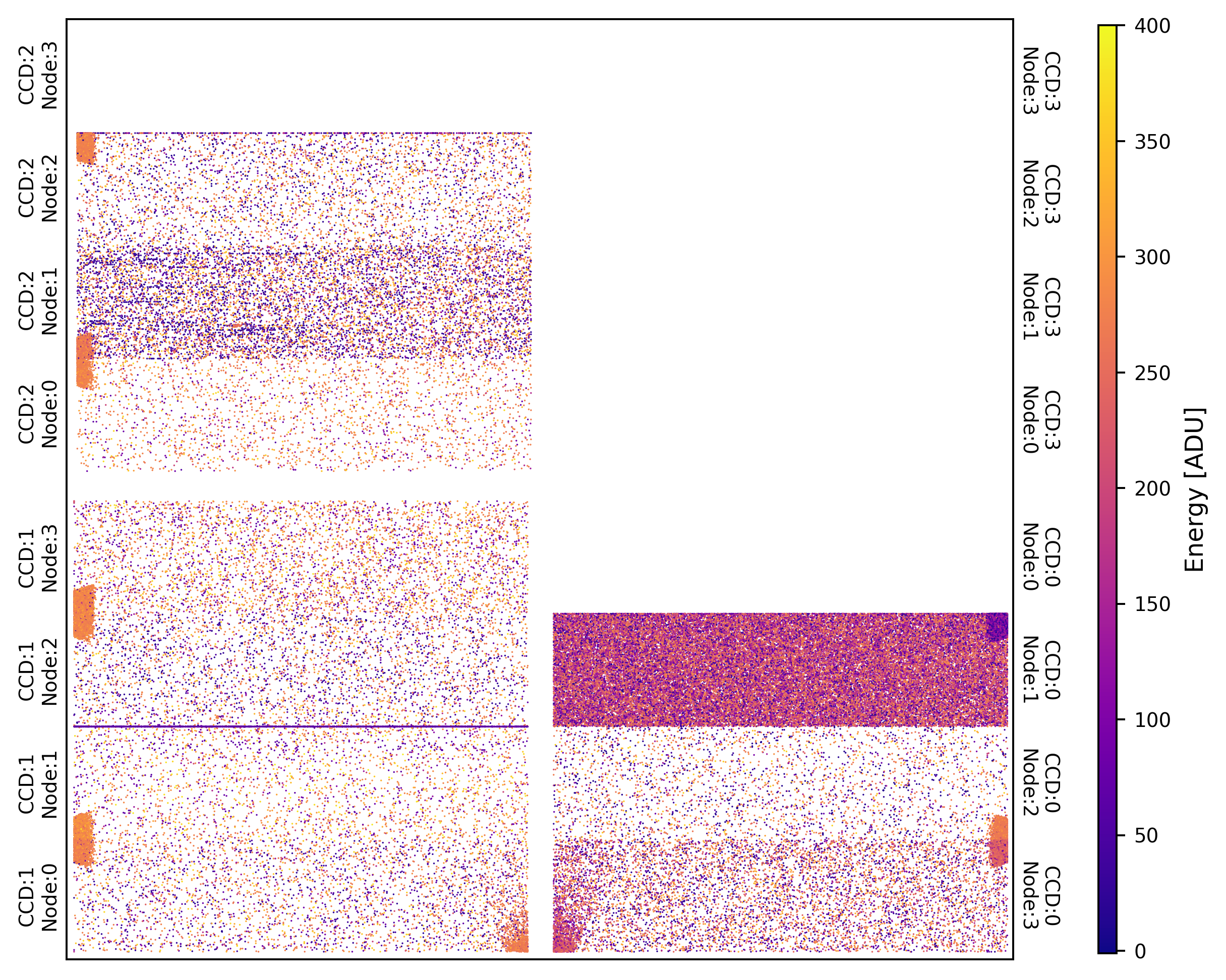}
	\caption{Data from the REXIS internal background calibration, performed in flight July 13 2018. Each putative x-ray event is marked by a colored circle. The illuminated regions from the \Fe sources are visible around the perimeter.  The colorscale is in units of uncalibrated energy, and each detector node has a slightly different gain, which causes the 5.9~keV photons to be recorded at slightly different raw energy values in each node. Some data quality issues are immediately obvious, for example the hot column in CCD~1 Node~1 and excess noise in CCD~0 Node~1 and CCD~2 Node~1.}
	\label{f:Fe_evt_dist}
\end{SCfigure}

REXIS carried eleven \Fe sources with a combined radioactivity of $\sim$~2~$\mu$Ci. As shown in Figure 2, these sources were located around the perimeter of the CCD array to enable monitoring each of 16 nodes in the 2 $\times$ 2 CCD array \citep{Masterson18}. 
One of the sources was placed on the inner side of the radiation cover, which was pasted and cured with a \Fe solution of 200~nCi (Figure~\ref{f:instrument}). The radiation cover source enabled the calibration of the entire detector array when the cover was closed and became invisible to the detector array after cover opening.

Figure~\ref{f:Fe_evt_dist} shows an example image produced by the full detector plane during the in-flight Internal Calibration Operation. Eight sources, each with 200~nCi, were arranged so that small regions ($\sim$~100 $\times$ 50 pixels) next to the readout buffer in each of 16 CCD nodes (two nodes per source) were exposed to \Fe X-rays. Two other sources (75~nCi each) were positioned so that a small region ($\sim$ 150 $\times$ 120 pixels) in the far side of the readout chain on each of the four nodes on both sides is illuminated. The ten \Fe sources on the DAM enable continuous monitoring of the spectral response of all 16 nodes with relatively low radioactivities, minimizing the area of potential contamination during X-ray observations of Bennu. In principle, charge transfer inefficiency (CTI) during readout could be monitored in four nodes through comparison of the spectral line resolution at the near- and far-end of the node; e.g., see markings {\it A} and {\it B} in Figure~\ref{f:Fe_sources} for CCD 1 node 0.

\subsubsection*{Energy Calibration}

\begin{SCfigure}
	%\centering
	\small
		\includegraphics[width=3.2in]{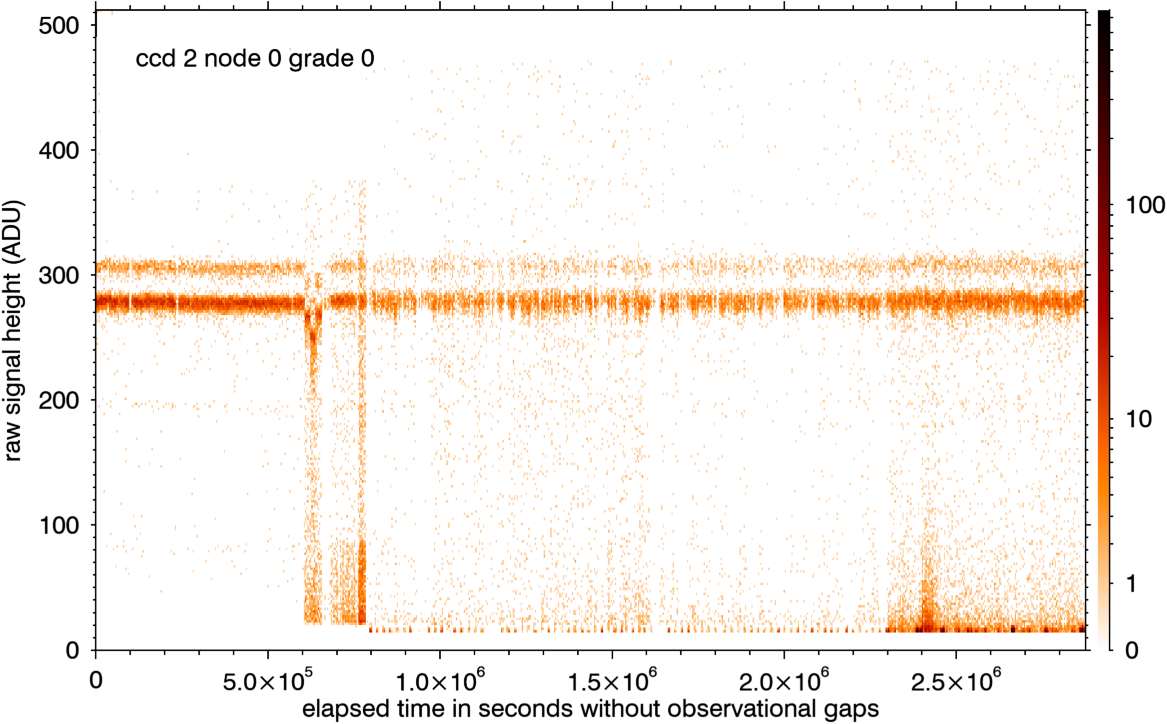}
	\caption{Raw pulse heights of valid single pixel events of CCD 2 Node 0 over elapsed time without observational gaps. The two horizontal streaks around 300 ADUs are 5.9 and 6.5~KeV Mn K-lines from the \Fe sources. The variations in these streaks indicate changes in either spectral gain, offset or both. The analysis showed the main cause is the shift in the spectral offset.}
	\label{f:Fe_pha}
\end{SCfigure}

The energy of each REXIS CCD event was calibrated based on the observed raw signal amplitude of the X-rays from onboard \Fe sources. Figure~\ref{f:Fe_pha} plots the raw signal heights of valid single-pixel X-ray events (Grade 0 events) in CCD~2 Node~0 over continuously elapsed time in seconds (i.e. without observational gaps). The two horizontal streaks around 300~ADUs are due to the 5.9 and 6.5~KeV Mn-K lines from the \Fe sources, which are more or less stable except for a few occasions (Section~\ref{sec:lightleak}). 

Assuming a linear relation between the raw signal ($A$) and the input X-ray energy ($E$) (measured in units of keV), at least two X-ray lines of known energies are required to constrain both spectral gain ($g$) and offset ($c$) parameters. Since the 6.5~keV line is much fainter than the 5.9~keV line, the former is not suitable for calibration of short term variations. A solution to this problem is to utilize the charge split events of the same energy over two pixels.
The expected raw signal of a single pixel event ($A_s$) or a double pixel event ($A_d$) from the Mn-\Ka line is given as    
\begin{eqnarray*}
    & A_s  = g E_{\mbox{\scriptsize 5.9 keV}} + c, &  \text{\ and} \\
    & A_d  = g E_{\mbox{\scriptsize 5.9 keV}} + 2 c. & 
\end{eqnarray*}

\begin{SCfigure}
%	\centering
    \small
		\includegraphics[width=3.2in]{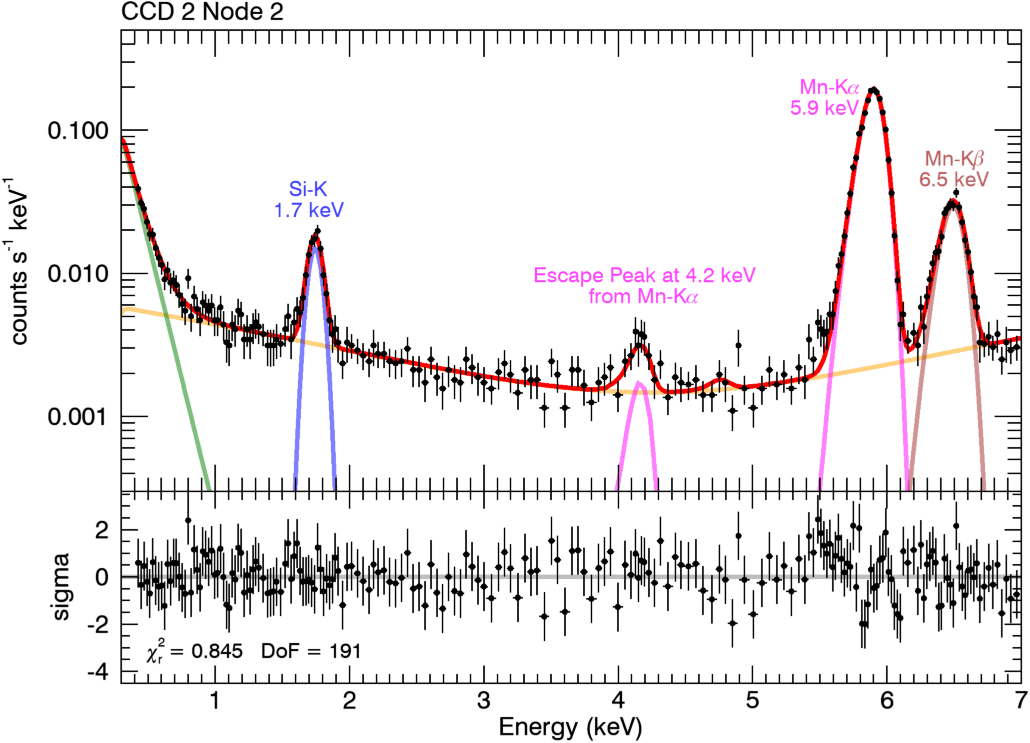}
	\caption{Energy calibrated X-ray spectrum of CCD 2 Node 2 during the Internal Calibration Operation. The \Fe source generates 5.9 and 6.5~keV Mn-K lines, and their escape peaks. The 1.7 keV Si-K line is generated by internal fluorescence from cosmic rays. The internal background is well-fit by a 2nd order polynomial function with a low energy exponential tail component (see Section~\ref{sec:excess} for more detail).}
	\label{f:Fe_spectra}
\end{SCfigure}

Figure~\ref{f:Fe_spectra} shows an example of a calibrated REXIS X-ray spectrum. These data were collected during the Internal Calibration Operation prior to the opening of the radiation cover, to enable the characterization of the internal background spectrum of each node as well as providing a snapshot of the spectral gain and offset. Even with the radiation cover closed, multiple X-ray lines beside the two Mn-K lines are present in the spectrum. These lines include the Si-K line at 1.7~keV, which is mainly caused by cosmic ray interactions in the silicon layer of the CCDs, and an escape peak at 4.2~keV, which is created by escaping Si-K lines fluoresced by the Mn-\Ka line from the sources. The escape peak at 4.8~keV due to the Mn-\Kb line is too faint to detect. 

\subsubsection*{Charge Transfer (In)Efficiency}

The difference in the peak value of the \mbox{Mn-\Ka} line between the near and far end of each node indicates a loss of the total charge due to the charge transfer across the node during readouts (e.g., markings {\it A} and {\it B} in Figure~\ref{f:Fe_sources}). During the Internal Calibration Operation, the charge transfer loss was measured to range from 0.6\% to 1.1\%, depending on the nodes, which is equivalent to 35-65~eV for the Mn-\Ka line. In the case of CCD 0 Node 1, the additional \Fe source on the DAM enables the charge transfer loss measurement even after opening the radiation cover. During the Orbital B observations, the charge transfer loss in CCD 1 Node 0 stayed about the same (approximately 1\%) as in the Internal Calibration Operation.

%%\subsubsection*{Changes in Q.E.}
\subsubsection*{Changes in Quantum Efficiency}

\begin{SCfigure}
	%\centering
	\small
		\includegraphics[width=3.2in]{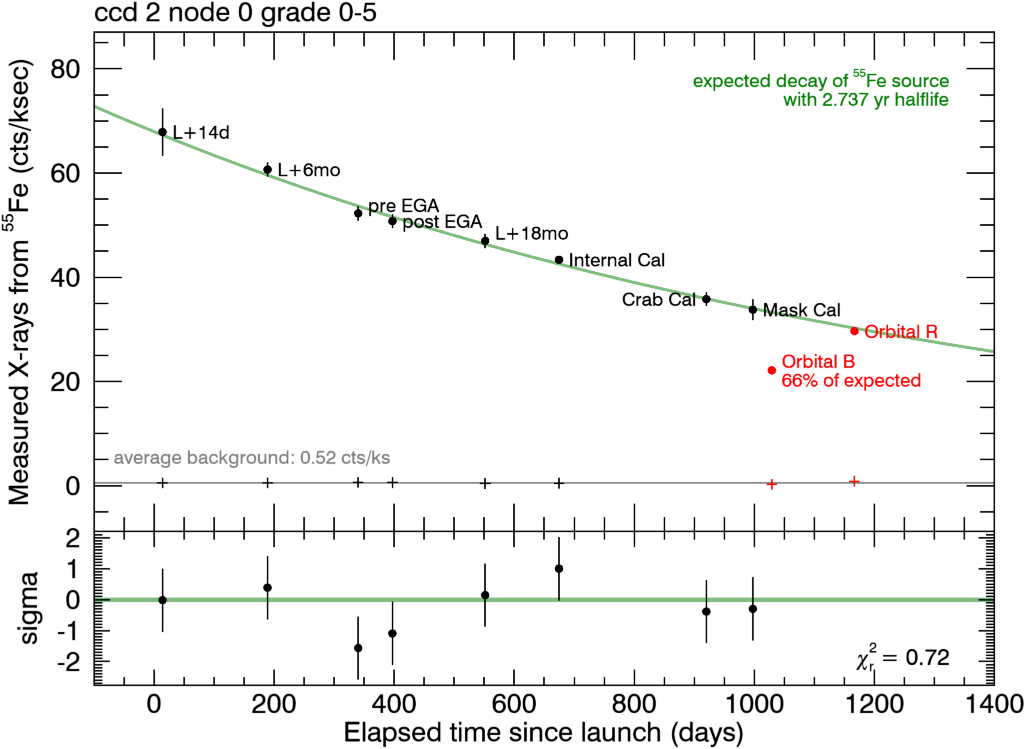}
	\caption{Measured flux changes against expected decay rate of the perimeter \Fe source in CCD 2 Node 0. During Orbital B, the measured flux rate dropped below the expected rate, indicating 34\% loss in the Q.E. The loss was due to a light leak, which induced a background of low energy events in the CCDs and spoiled the event grading algorithm (causing single-pixel events to appear as multi-pixel events). The measured background rates (crosses) in the 5 - 7 keV band remained steady. The observed flux returned to the expected rate during Orbital R, following a change to the CCD split-charge threshold and greater orbital altitude.}
	\label{f:Fe_intensity}
\end{SCfigure}

In addition to monitoring the spectral response of the CCDs, the \Fe sources enable monitoring of relative changes in the quantum efficiency of each CCD node. With a half life of 2.7~years, the X-ray flux from the \Fe sources decreased by more than a factor of two over the timescale of REXIS operations. Figure~\ref{f:Fe_intensity} shows the observed flux intensity of the two Mn-K lines from the \Fe sources in CCD 2 Node 0. The observed flux decayed as expected until Orbital B, when there was a significant drop. The decrease in the \Fe detection rate during Orbital B was ultimately traced to the presence of optical light in the detector; see Section~\ref{sec:lightleak}. Optical photons from the light lead induced a background of low-charge events in the CCDs, which spoiled the event grading algorithm (causing single-pixel events to appear as multi-pixel events).

%that raised the number of low energy background events, drastically reducing the ability of the on-board software to recognize and count single pixel (Grade 0) X-ray events. The REXIS CCDs are sensitive to optical and UV wavelengths, and thus are equipped with 320~nm Al OBF directly deposited on the CCD silicon substrate to shield optical/UV lights. While the OBF performed as expected according to the analysis of the OBF Calibration data, light leaks did occur along the {\it uncoated} side and edge of the CCDs, which was evident due to the higher low energy events near the CCD edges. See Section~\ref{sec:lightleak} for more details.

\subsection{Spectral Resolution and Background Rates} \label{s:response}

\begin{table*}
\small
\caption{Spectral Response and Internal Background Rate of Five Prime CCD Nodes in REXIS}
\centering
\begin{tabular}{ @{\extracolsep{4pt}}ccccccc }
\hline\hline
CCD  & Node & \multicolumn{3}{c}{Spectral Resolution$^a$}     & \multicolumn{2}{c}{Background Rates$^b$} \\
\cline{3-5}\cline{6-7}
ID   & ID   & Region$^c$ & Int. Cal. & Orbital B Obs.           & Int. Cal. & Orbital B Obs.   \\
\hline
0   & 2     & near      & 153$^{+26}_{-22}$ eV  & 172$^{+22}_{-22}$ eV & 1.53(6) & 3.08(4)\\
\multirow{2}{*}{1}      & \multirow{2}{*}{0}      
            & near      & 247$^{+25}_{-27}$ eV  & 282$^{+23}_{-23}$ eV & \multirow{2}{*}{1.61(6)} & \multirow{2}{*}{5.95(6)} \\
    &       & far       & 294$^{+26}_{-25}$ eV  & 304$^{+24}_{-24}$ eV &\\
1   & 2     & near      & 230$^{+24}_{-25}$ eV  & 292$^{+22}_{-22}$ eV & 1.63(6) & 4.28(5)\\
2   & 0     & near      & 153$^{+23}_{-22}$ eV  & 215$^{+22}_{-22}$ eV & 1.59(7) & 1.95(3)\\
2   & 2     & near      & 159$^{+23}_{-23}$ eV  & 213$^{+22}_{-22}$ eV & 1.90(6) & 3.58(4)\\
\hline
\multicolumn{7}{l}{$^a$FWHM at 5.9 keV from the perimeter source regions}\\
\multicolumn{7}{l}{$^b$counts ks$^{-1}$ cm$^{-2}$ keV$^{-1}$ in the 2 -- 3.5 keV band of the full node}\\
\multicolumn{7}{l}{$^c$Illumination region: near (marked {\it A} for CCD 1 Node 0 in Figure ~\ref{f:Fe_sources}) or far (marked {\it B}) }\\
\multicolumn{7}{l}{end of the node w.r.t. the framestore.}
\end{tabular}
\label{t:resolution}
\end{table*}

Table~\ref{t:resolution} summarizes the spectral resolution and background rates of the five stable CCD nodes in the REXIS detector plane. Interestingly, all five of the most stable nodes have an even number node ID (Figure~\ref{f:Fe_sources}). This is not a coincidence, and was later identified in part due to the interference from the residual signal in CCD~3 which suffered a severed flexprint cable, since the location of the readout electronics in the MEB for the nodes with an odd number ID is more susceptible to the interference. For more details, see \citet{Masterson18}. The spectral resolutions of the five prime CCD nodes in Table~\ref{t:resolution} range from $\sim$150 to 300 eV in FWHM. 
The nodes in CCDs 0 and 2 show finer spectral resolutions than those in CCD 1. This is interesting since CCD 1 was considered as Grade A while CCDs 0 and 2 were Grade C before the assembly, and we suspect that the performance degradation in CCD 1 is also related to the interference from the power line in CCD 3. 

Comparing the measurements during the Internal Calibration Operation in July 2018 and the Orbital B Observations in July 2019, there is a noticeable degradation in the CCD spectral resolution. Assuming a quadratic contribution of each noise component, the additional degradation ranges from 80 to 180~eV, which is likely due to radiation damage during the 10-month period between opening of the radiation cover and Orbital B. In CCD~1 Node~0, there are two regions illuminated by the \Fe sources, and the degradation in the spectral resolution at the far-end of the node (w.r.t.~the readout buffer: labeled as {\it B} in Figure~\ref{f:Fe_sources}) appears relatively small, but the relatively poor resolution of the node makes it hard to isolate the effect of radiation damage during the charge transfer. 

The background rates in Table~\ref{t:resolution} indicate the average count rate in the 2 to 3.5~keV band, which is free of instrumental spectral lines (see Figure~\ref{f:Fe_spectra}). The background rates during the Orbital B observations include the contributions from the Cosmic X-ray Background (CXB), potential X-ray fluorescence (XRF) from Bennu, and the light-leak induced noise. CCD~2 Node~0 shows the smallest increase in the background rate during the Orbital B observations compared to the Internal Calibration Operation. This is consistent with the CXB being more occulted by Bennu in the spatial region imaged by CCD~2 Node~0 (see \cite{Hoak21}). For most other nodes, the spatial orientation of the asteroid during Orbital B was such that the CXB dominates the observed X-ray flux.

\subsection{Response Matrix Function (RMF) Modeling using Onboard \Fe Sources and Simulations} \label{sec:rmf}

The response of an X-ray sensor to incoming X-rays is commonly described by the Response Matrix Function (RMF) and the Auxiliary Response Function (ARF). The RMF is a collection of  probability distributions as a function of raw pulse height for the range of X-ray energies to which the sensor is sensitive. Thus, the RMF essentially enables the translation between input X-ray energies and possible corresponding detection signal heights. The ARF describes the detection efficiency in units of effective area as a function of X-ray energy. The ARF involves the Q.E. of the sensor and other relevant factors (e.g., absorption loss due to the optical blocking filter (OBF) in REXIS). Combining the RMF and ARF, one can fully characterize how often a given X-ray photon would interact with the sensor and how much signal the input X-ray would induce in the sensor when it does.

The REXIS team had planned an extensive pre-flight ground calibration program for the REXIS CCDs to establish both the RMF and the ARF for each node, but this program had to be abandoned due to schedule delays during the assembly. Instead, the RMFs for the REXIS CCDs were calculated based on X-ray CCD response simulations and spectral measurements of the onboard \Fe calibration sources during in-flight operations. Similarly, the ARFs were estimated based on the observations of the Crab Nebula, a bright celestial X-ray source with a steady X-ray flux and spectrum as described in Section \ref{sec:arf}.  

Since the onboard \Fe sources produce relatively high energy Mn-K lines at 5.9 and 6.5 keVs, the CCD response at low energies has to be modeled through X-ray CCD simulations. Note that the Si-K lines at 1.7 keV observed in the REXIS CCDs (e.g., Figure~\ref{f:Fe_spectra}) have a  different response compared to 1.7 keV X-rays that are incident on the CCDs, since the former is generated within the silicon substrate inside the CCDs. The X-ray response simulations were performed using a simulation tool developed for the X-ray imaging Spectrometer (XIS) on-board the {\it Suzaku} mission \citep{Prigozhin98, Prigozhin00, Prigozhin03}. 

\subsubsection*{XIS Simulator}
The XIS simulator performs numerical modeling of back-illuminated (BI) CCDs using Monte-Carlo techniques. Simulated photons of a given energy interact with a model of the surface of the CCD at random locations, and the behavior of each photon is evolved using basic physics of photon interactions with silicon. The program determines the site of photon interaction with silicon atoms, based on the probability of absorption. The exposed surface of the BI CCD is modeled with a uniform layer of SiO$_2$. The simulator determines whether fluorescence took place; if the answer is yes, the fluorescent photon is emitted in a random direction and traced, either escaping the CCD or re-interacting with the silicon. In the case of a re-interaction the fluoresced photon is treated with the same interaction modeling as the original photon. 

Next, the simulator models the creation and expansion of the charge cloud and charge losses due to interaction with the surface. The program models the distribution of charge across the pixels surrounding the photon interaction site, simulates the charge transfer from the imaging section into the frame store, from the frame store into the serial register, and from the serial register into the readout node.  Readout noise is added to each pixel (to simulate noise in the electronics chain), and a standard 3 $\times$ 3 pixel event is written to the output event list through an event finding algorithm. See \citet{Prigozhin00} for more details.

\begin{SCfigure}
%	\centering
    \small
	\includegraphics[width=3.2in]{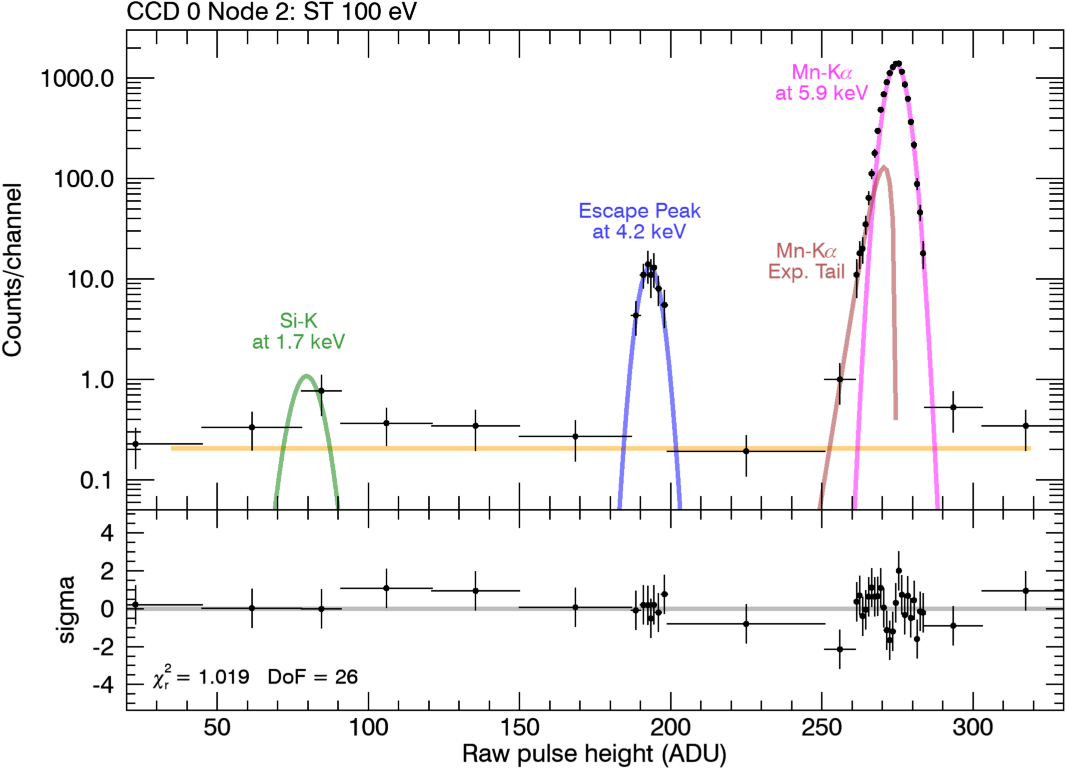}
	\caption{Simulated X-ray spectrum designed to produce the single-pixel spectral response in CCD 0 Node 2 for 5.9 keV Mn-\Ka X-rays. The split charge threshold of 100 eV and two readout noise components of 6.5 e$^-$ and 11.5 e$^-$ are used for the spectral response. The main peak for 5.9 keV X-rays in the spectrum is well described by a Gaussian line (pink) and an exponential tail component (brown): see Section~\ref{s:linemodel} and \citet{Prigozhin00}.}
	\label{f:XIS_sim_spectra}
\end{SCfigure}

Some of the key simulation parameters in modeling the observed \Fe X-ray spectra are the thickness of the silicon substrate layer, electronics readout noise, and the split charge threshold. The thickness of the silicon substrate layer (45 $\mu$m for the REXIS CCDs) determines how wide the charge cloud can spread while it drifts down from the interaction site. The split charge threshold (ST), which can be set for REXIS by an uplink command, is used to determine if the given CCD event is of a single pixel or multi-pixel origin. It ranges from about 100 eV to a few 100s~eV, depending on the node.

The electronics readout noise describes the noise characteristics of the readout chain. REXIS occasionally downlinked a full image of the charge distribution over the entire detector plane collected during one readout cycle (4 sec) for diagnostic purposes. These full frame images indicated that two readout noise components of high and low frequencies were present in the REXIS CCDs. We incorporated these two noise components in to the RMF calculation by combining the results of two sets of the XIS simulations with two separate readout noise parameters.

Figure~\ref{f:XIS_sim_spectra} shows the simulated spectrum for 5.9 keV X-rays, where the Si-K and escape lines are also present. We tuned the model parameters to closely match the probability distribution of 5.9 keV X-ray photons to the observed raw signal height distribution of the Mn-\Ka line at 5.9 keV from the on-board \Fe sources during the Internal Calibration Operation. The full RMF was constructed using this configuration of XIS simulations for the rest of the input X-ray energies. We repeat this process separately for single (Grade 0) and double (Grades 2-5) pixel events for each REXIS CCD node.

\begin{SCfigure}
%	\centering
    \small
		\includegraphics[width=3.2in]{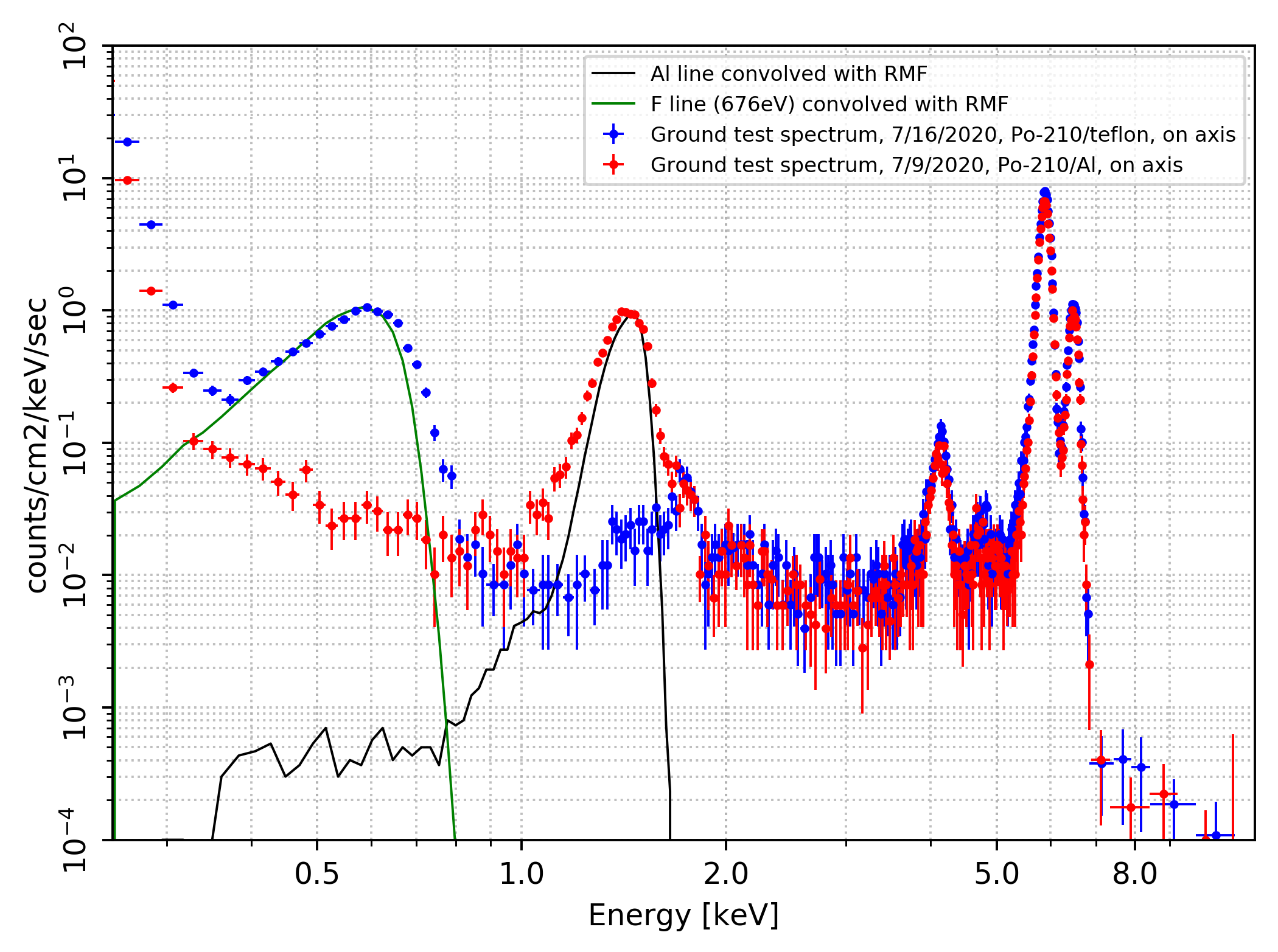}
	\caption{X-ray spectra of low energy X-ray lines taken with the flight spare CCD on the ground.  A \Po source was used to illuminate a Teflon or aluminum target, which produced the 0.67 keV F-K (blue) and 1.5 keV Al-K (red) lines, respectively. An \Fe source was present in both cases. The solid lines show the modeled detector response using the RMF of CCD 2 Node 2 for the flight instrument.}
	\label{f:lowEnergyLine}
\end{SCfigure}

\subsubsection*{Applying RMFs}
The assembled RMFs were used for the spectral fit in Figure~\ref{f:Fe_spectra}, and they correctly modeled the asymmetric spectral shape of the Mn-K lines and their escape peaks. The XIS simulations also predict the Si-K line from simulated input X-rays, but the observed Si-K line has a much greater amplitude, and is likely triggered by cosmic rays rather than incoming X-rays. In our model of the REXIS CCD internal background noise, we used an additional Gaussian line to properly described the Si-K line in the observed spectra. 

As mentioned in Section~\ref{s:response}, the spectral resolution varies spatially in each node and is expected to degrade with time. In order to capture small changes in spectral resolution with our fixed RMF, individual spectral lines (including the Mn-K lines) are modeled using a Gaussian function with a narrow but nonzero width, rather than with a delta-function. The full-width-half-maximum (FWHM) of the Gaussian function is allowed to vary. The spectral resolution in Table~\ref{t:resolution} for a given node is modeled by the same RMF, but the the width of the observed Mn-K lines is determined by the quadratic sum of two components: the width of the Gaussian function (which is a free parameter, in a narrow range) and the width set by the RMF (which is fixed in the model for the same node for a given energy).

\subsubsection*{Validation of RMFs at Low Energies}

The low energy performance of the constructed RMFs was indirectly validated by the data taken with a flight spare CCD on the ground. We used a \Po $\alpha$ source to excite targets of fluorine (Teflon) and aluminum and observed the CCD response of the F-K and the Al-K fluorescent lines. In Figure~\ref{f:lowEnergyLine}, we compare the observed X-ray spectra of the 0.67 keV F-K and 1.5 keV Al-K lines from the spare CCD with the response models in the RMF of CCD 2 Node 2 for the lines. Even without fine tuning the RMF to the performance of the spare CCD used to collect the data, the agreement between the data and the model is good.

\subsection{Auxiliary Response Function (ARF) Modeling using Observations of the Crab Nebula and Sco X-1} \label{sec:arf}

The onboard \Fe sources enabled monitoring of relative changes in the quantum efficiency of the CCDs, but it is difficult to estimate the absolute efficiency of the CCDs (and thus the effective area of REXIS) with the \Fe sources due to the limited energy coverage of the spectral lines and the complex geometric effects. In order to measure the effective area of the CCDs as a function of incident energy, we observed the Crab Nebula, a bright, stable X-ray source with a precisely measured flux and spectrum \citep{Weisskopf10}. REXIS observed the Crab Nebula on March 16, 17, 23, and 24 in 2019 for a total of 16 hours.  Since the Crab Nebula appears as a point-like source to REXIS with $\sim$ 26$'$ angular resolution, only about 41\% of the CCD pixels were exposed to the Crab at a given time because of the REXIS coded-aperture mask. To ensure approximately equal exposure for all detector pixels, \OSIRIS oriented the REXIS boresight to eight different points on the sky for each day of the operation; the pattern of observation targets is shown in Figure \ref{f:crab_pointings}. During the second two days (March 23-24), we observed a spatially uniform time dependent background in the CCD data, which we attribute to increased solar particle flux associated with a passing heliospheric current sheet (Parker spiral). Excluding the data with elevated background leaves about 8 hours of useful integration time, which provided sufficient CCD counts to measure the energy-dependent effective area for the individual CCD nodes.

\begin{SCfigure}
	%\centering
	\small
		\includegraphics[width=3.2in]{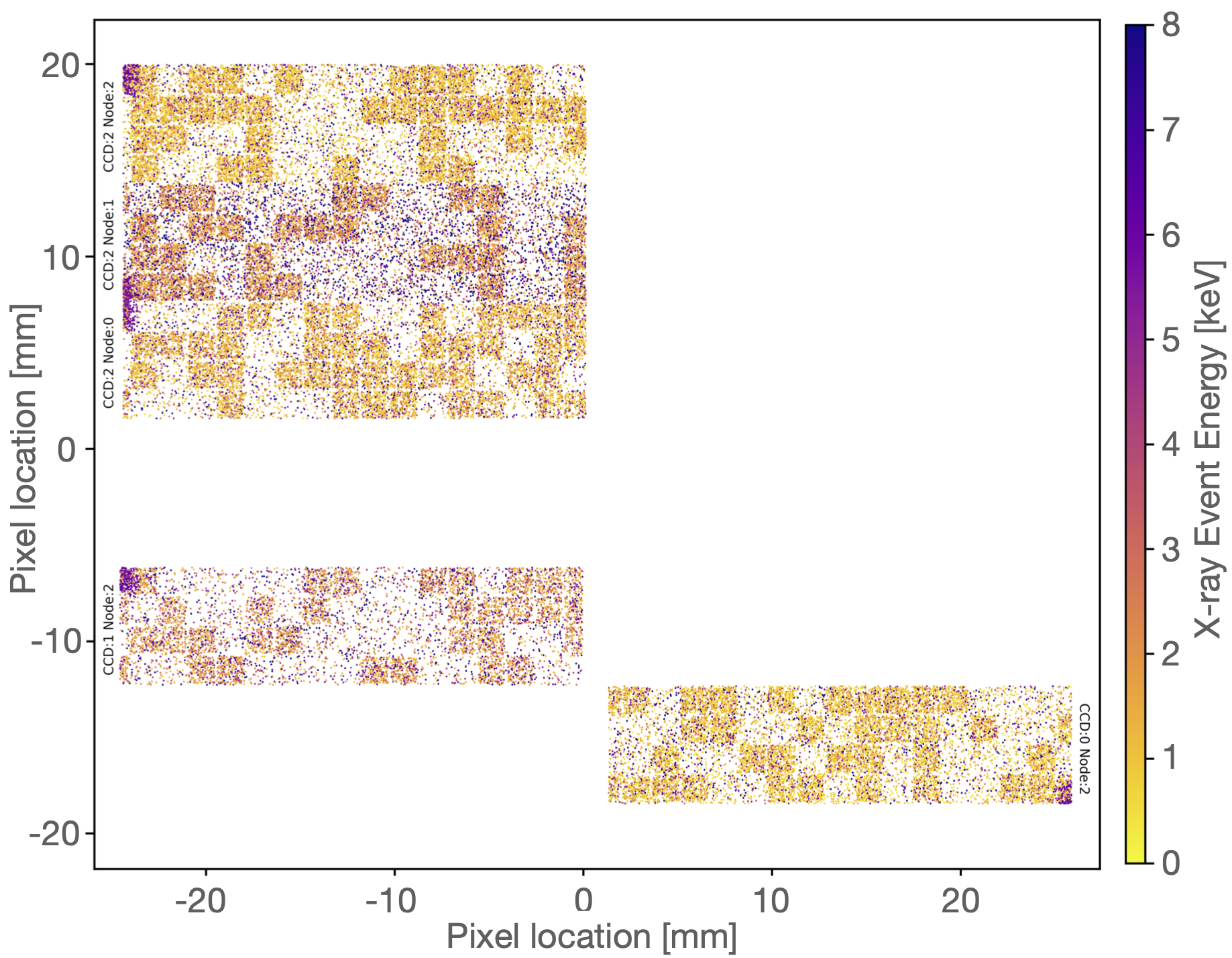}
	\caption{Data from the REXIS observations of the Crab Nebula, March 16-17 2019. Only five detector nodes were enabled for this observation. The pattern of the REXIS coded aperture mask is clearly visible. The data in this figure are collected from about one hour of exposure time.}
	\label{f:crab_detplane}
\end{SCfigure}

We calibrated the energy values of the CCD event-list data using the on-board \Fe sources as described in Section~\ref{sec:Fesources}. To separate the X-ray signal of the point-like Crab Nebula from the diffuse cosmic X-ray background (CXB), we separated the data from each pointing into events from pixels that were exposed to the Crab Nebula (the ``on-source" pixels) and events from pixels that were not exposed to the Crab (the ``off-source" pixels). For each pointing, we subtracted the area-weighted off-source spectrum from the on-source spectrum. The resulting spectrum (on-source minus off-source) contains only the spectrum of X-rays from the Crab Nebula. We used this observed spectrum, together with the known Crab X-ray spectrum and our model for the detector response (see Section~\ref{sec:rmf}) to measure our effective area as a function of X-ray energy. The analysis was performed twice for each CCD node, once for single-pixel events (Grade 0) and again for double-pixel events (Grades 2-5). Higher order multi-pixel events were not modeled due to their high background noise and poor energy resolution.

\begin{SCfigure}
	%\centering
	\small
		\includegraphics[width=3.2in]{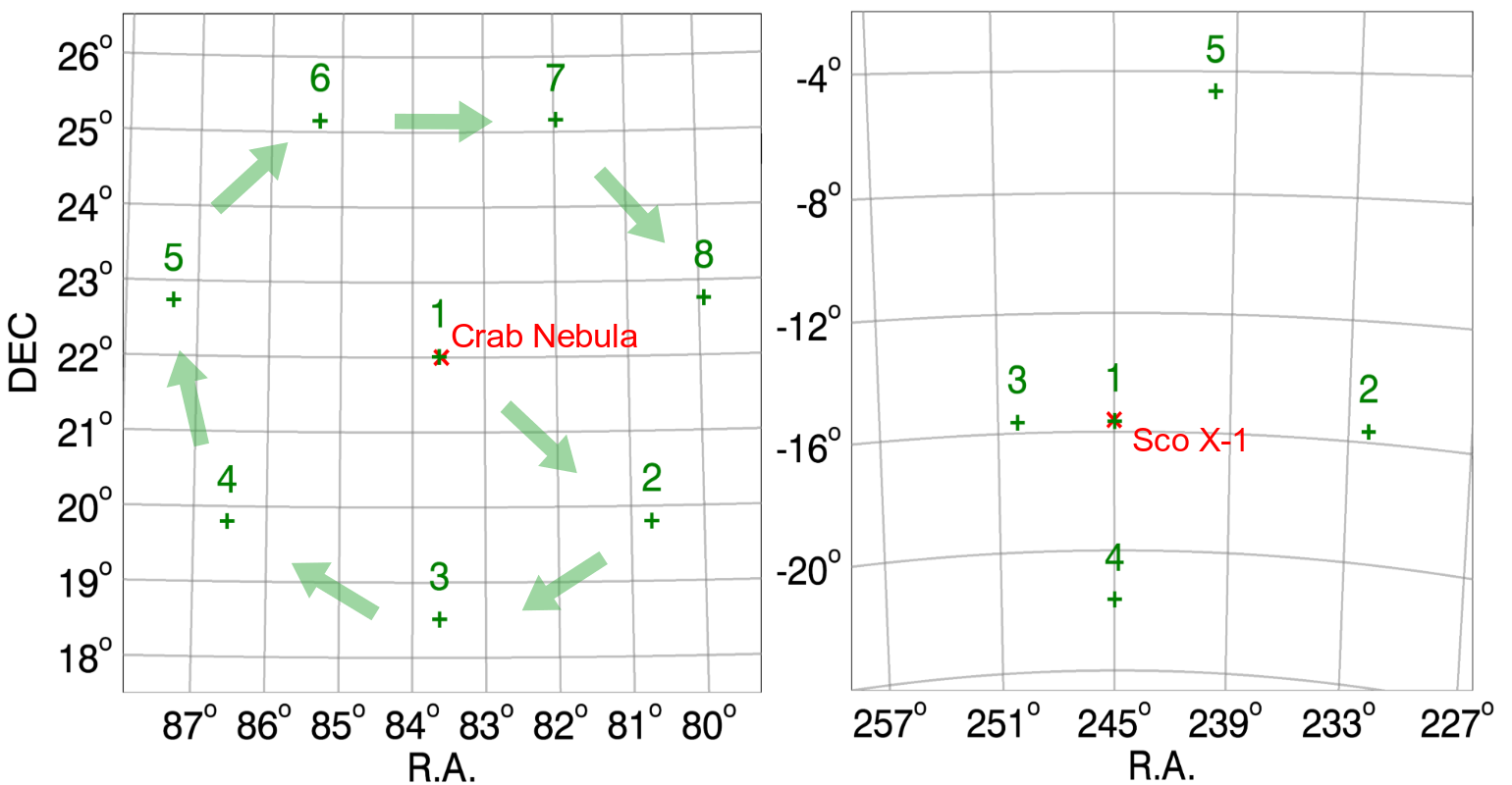}
	\caption{REXIS pointing sequences during the Crab Nebula calibration in March 2019 and the Sco X-1 observations for Mask Calibration Operation in June 2019.  
	See Tables~\ref{table:crab_observations} for pointing coordinates.}
	\label{f:crab_pointings}
\end{SCfigure}

\begin{SCfigure}
	%\centering
	\small
		\includegraphics[width=3.2in]{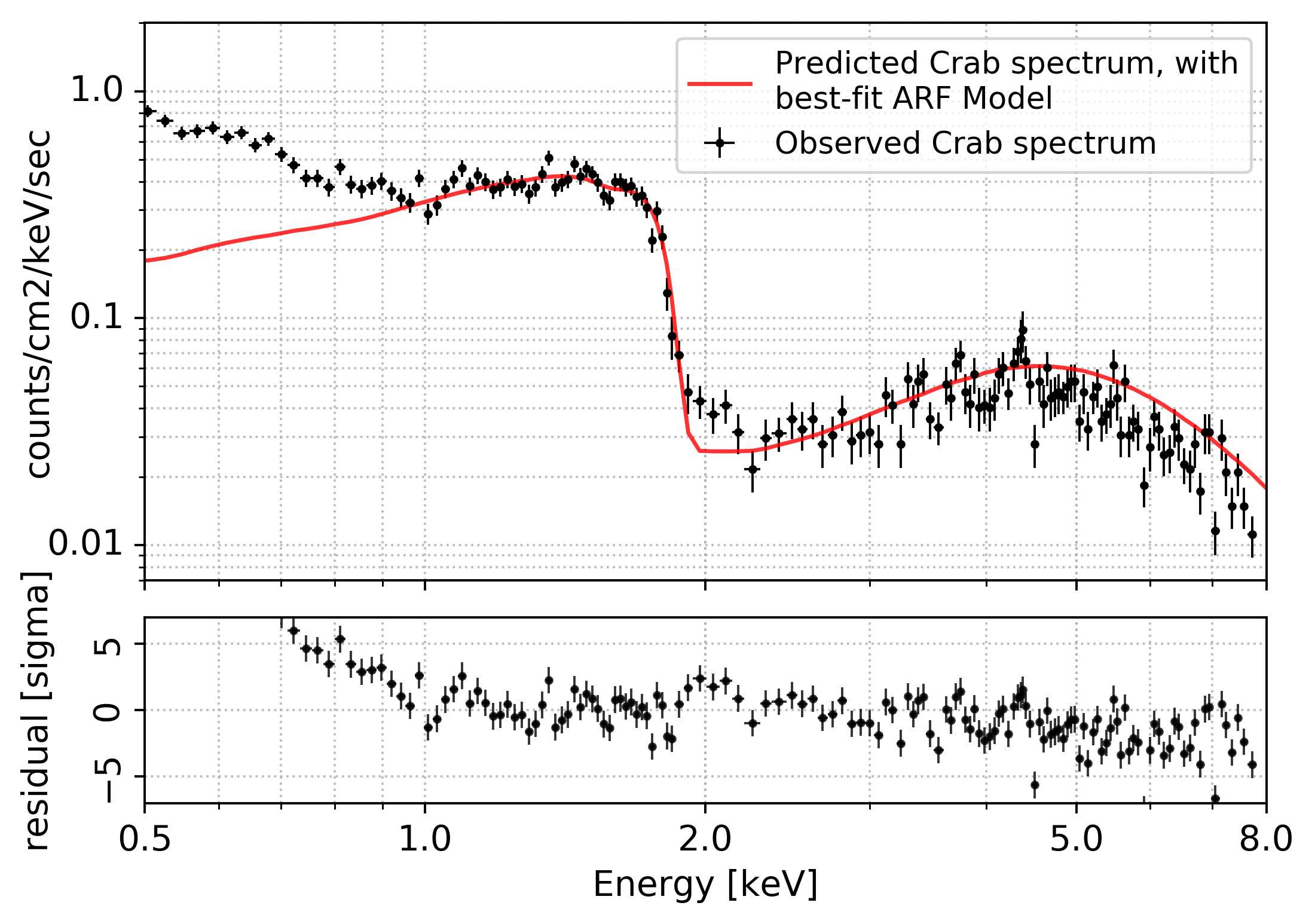}
	\caption{Results from the fit of the CCD 2 Node 0 ARF model for single-pixel events to the Crab Calibration data. The reduced $\chi^2$ of the fit is 1.55 between 1 and 4~keV.}
	\label{fig:ccd20_crab_fit}
\end{SCfigure}

\begin{SCfigure}
	%\centering
	\small
		\includegraphics[width=3.2in]{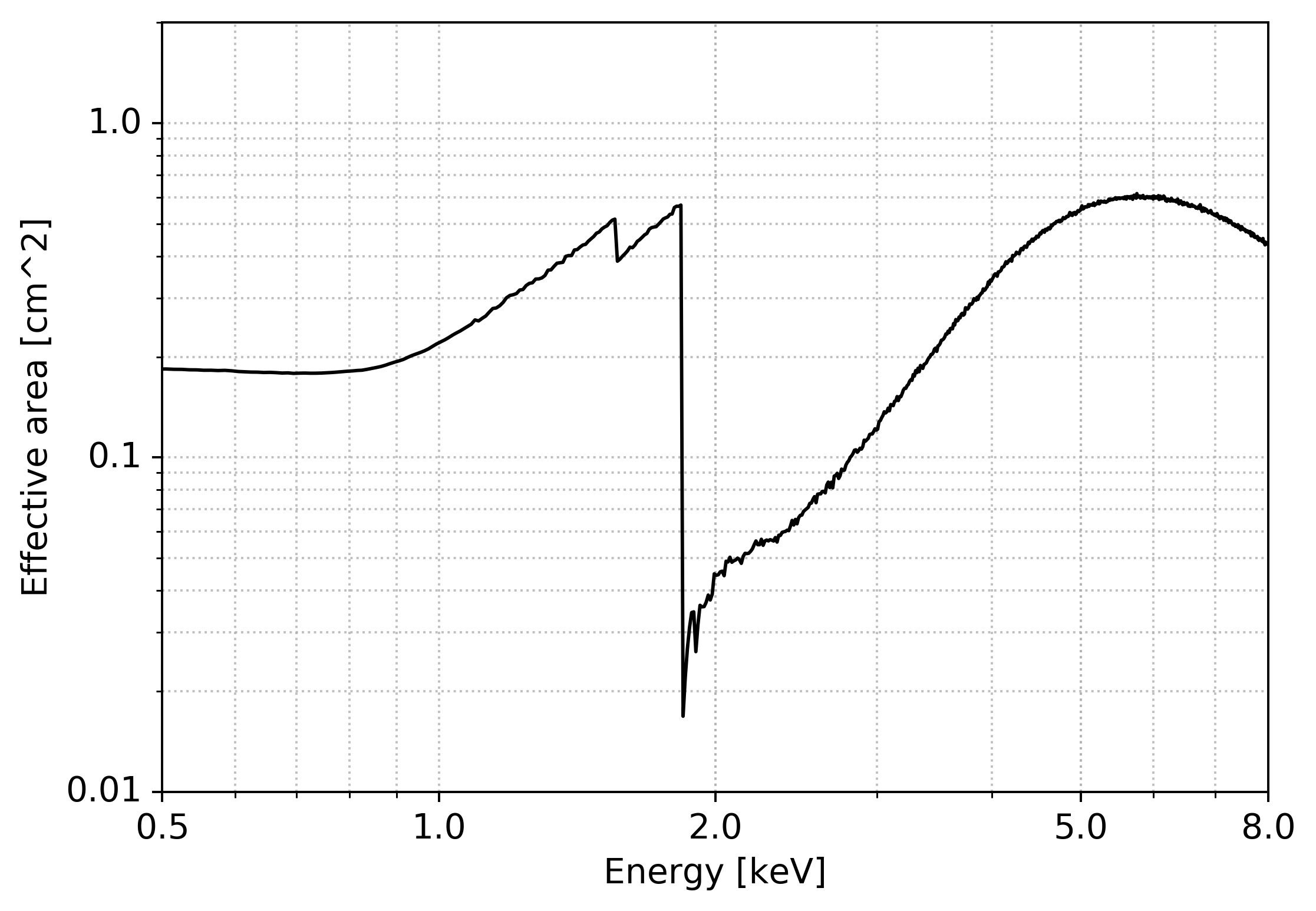}
	\caption{REXIS CCD 2 Node 0 effective area, for single-pixel (Grade 0) events. The node has a physical area of 1.509~cm$^2$, implying a single-pixel detection efficiency of 22\% at 1.25~keV.}
	\label{fig:ccd20_arf}
\end{SCfigure}

To measure the effective area of the CCDs, we used the Sherpa Python package \citep{Freeman01} to convolve the modeled X-ray spectrum of the Crab Nebula from \citet{Weisskopf10} with the RMF as described in Section~\ref{sec:rmf} and a parametric ARF model, and searched for the ARF parameter values that gave the minimum $\chi^2$ between the background-subtracted observed spectrum and the response-convolved model spectrum. The free parameters in our ARF model were the thickness of the aluminum optical blocking filter, the detection efficiency due to split-charge events, and an overall normalization factor. 

Figure~\ref{fig:ccd20_crab_fit} shows the results of the fit to the observed spectrum for a single node (CCD 2 Node 0). Below 1~keV, there is an excess of events that does not fit the model; this is discussed in greater detail in Section~\ref{sec:excess}. Table~\ref{t:effective_area} summarizes the effective area of the prime nodes at particular elemental lines of interest. The five prime REXIS nodes collectively provide an effective area of 4.4 cm$^2$ at 1.25 keV. Figure~\ref{fig:ccd20_arf} shows the best-fit effective area model of CCD 2 Node 0 for single pixel events.

% Note: CCD10 was disabled for the Crab cal. Values are from ARFv3 except for CCD12, which had a poor fit; that node uses ARFv2
\begin{table*}
\centering
\small
\caption{REXIS effective area at elemental line energies for single-pixel events (effective area of double-pixel events in parenthesis). Values are for a full detector node, and do not include the throughput of the coded aperture mask (typically about 41\%). The physical area of a node is 1.509~cm$^2$.}
\begin{tabular}{ c|c|c|c|c }
\hline\hline
CCD  & Node & \multicolumn{3}{c}{Effective area [cm$^2$]} \\
ID   & ID   & Mg (1.25~keV) & Si (1.74~keV) & S (2.3~keV)   \\
\hline
0 & 2     &   0.33 (0.69) & 0.49 (0.65) & 0.06 (0.29) \\
1 & 2     &   0.32 (0.13) & 0.29 (0.42) & 0.08 (0.27) \\
2 & 0     &   0.33 (0.80) & 0.50 (0.76) & 0.06 (0.31) \\
2 & 2     &   0.44 (0.95) & 0.57 (0.89) & 0.10 (0.39) \\
\hline
\end{tabular}
\label{t:effective_area}
\end{table*}

We performed an additional calibration in June 2019, following spacecraft operations that were expected to incur significant temperature variations on the REXIS mask (see Section~\ref{sec:mask}). The REXIS team coordinated simultaneous observations of the bright variable X-ray source Sco X-1 with the \NICER experiment \citep{Gendreau16} onboard the International Space Station. The observation with REXIS lasted about 18~ksec with five separate offset pointings (see \ref{table:crab_observations}), and \NICER observed the source with a total exposure of 9~ksec in five orbits over the same period. The \NICER team provided a fit of the Sco X-1 spectrum with a thermal disk model, and we used that model as a fixed incident spectrum for our calibration. We reduced the Sco X-1 data in the same manner as our Crab Nebula calibration and fit the source spectrum derived from \NICER to our effective area model for each parameter setting, finding agreement with the \NICER data above 1~keV (see Figure~\ref{fig:spectra_LEE}).

%Furthermore, during the Sco X-1 observation, REXIS cycled through different settings of the split-charge threshold parameter, to measure the ARF at different parameter values. 

\subsection{Optical Light Sensitivity; Light Leak} \label{sec:lightleak}

The REXIS detector was designed with a novel optical blocking filter (OBF) to prevent visible wavelength (optical) photons from interacting with the CCDs. The REXIS OBF is a thin layer of aluminum, approximately 320~nm thick and directly deposited on the top surface of the four CCID-41 chips \citep{Thayer21}. A directly-deposited OBF on the array simplifies the overall detector assembly, which can otherwise be challenging for addressing all sources of direct or scattered light.  However, a directly deposited OBF can experience pinholes, or micron-sized areas with a significantly thinner deposition. CCD pixels beneath pinholes will have greater exposure to optical photons, which can generate sufficient charge to masquerade as X-ray events. For REXIS, with limited downlink capacity, the number of CCD pixels affected by pinholes could not be larger than a few hundred, otherwise events from these pixels could overwhelm the REXIS event-list data and cause true X-ray events to be discarded. To prevent the saturation of the CCD data, the REXIS flight software included an optional hot-pixel mask, i.e. a list of pixel addresses to ignore when generating the list of X-ray events. The hot pixel mask was uploaded to the REXIS firmware at each power-on cycle.

Since the complete number of pinholes was not measured before the launch, REXIS performed two observations of the sun-illuminated asteroid prior to the main observation campaign, to measure the number of likely pinholes and to confirm that the hot-pixel mask was performing as expected. During the L+30 months operation, in February 2019, REXIS collected three hours of event-list data, as well as five frames of the full detector area. Pixels that recorded significant charge in all of the frames were labeled as pinholes and included in the hot pixel mask. Ultimately, 1234 pixels (out of 1.5 million pixels in six detector nodes; less than 0.1 percent) were identified as bright and likely associated with pinholes, and were included in the hot pixel mask. REXIS successfully tested the hot pixel mask during the OBF Calibration in April 2019, and used the mask in all subsequent operations. During the primary asteroid-observation periods in Orbitals B and R, there was no evidence that pinholes contributed to the event list data.

While the OBF successfully attenuated optical photons from Bennu incident on the top surface of the REXIS CCDs, the CCDs still occasionally suffered from optical photon contamination. The inner surface of the REXIS XIS tower was coated with gold in order to minimize contamination from X-ray fluorescence generated internally within the instrument. However, this gold coating created a surface for scattering  optical photons inside of the tower. As a result, there was a path for sunlight incident on structures on the \OSIRIS instrument deck to scatter into the REXIS CCD tower and, via multiple internal reflections, arrive at the uncoated back side surfaces of the CCD substrate (see also Section~4.3 in \citet{Rizk18}). This path for scattered light was first observed following the opening of the REXIS radiation cover, when the CCDs experienced periods of large count rates correlated with times when the REXIS boresight was less than about 110$^{\circ}$ from the Sun. This was most noticeable during calibration operations in the second half of 2018 (e.g., CXB and 1st Crab Calibrations). As a result, the data from these operations was of limited use for instrument calibration. For later operations (starting with the 2nd Crab calibration), the REXIS commanding was adjusted to regenerate the CCD bias map at regular intervals, which mitigated the effect of diffuse scattered light. 

During asteroid observations in Orbital B, where the Sun-illuminated asteroid surface was directly in the REXIS field of view, there were further complications due to the light leak. Despite regenerating the bias map approximately hourly, REXIS experienced short periods (tens of minutes or less) when the CCD event list data was dominated by low-energy events, with a high density of counts near the readout edge of the CCDs where the flexprint cable connected to the CCD chip. Correlated with these periods, the rate of \Fe events was sharply reduced, by as much as 50\% in some CCD nodes. Ground testing and simulation \citep{Lambert20} demonstrated that optical photons from the asteroid were leaking into the CCDs along the readout edge, and these photons were generating low-energy events with sufficient rate to shift the grade of true single pixel X-ray events. During Orbital R, the onboard event-processing parameters were changed to accept more multi-pixel events, and the detection efficiency measured by the \Fe sources returned to nominal (see Figure~\ref{f:Fe_intensity}).

\subsection{Excessive X-ray Events at Low Energies below 1 keV} \label{sec:excess}

\begin{figure}
	\centering
	    \includegraphics[width=3.4in]{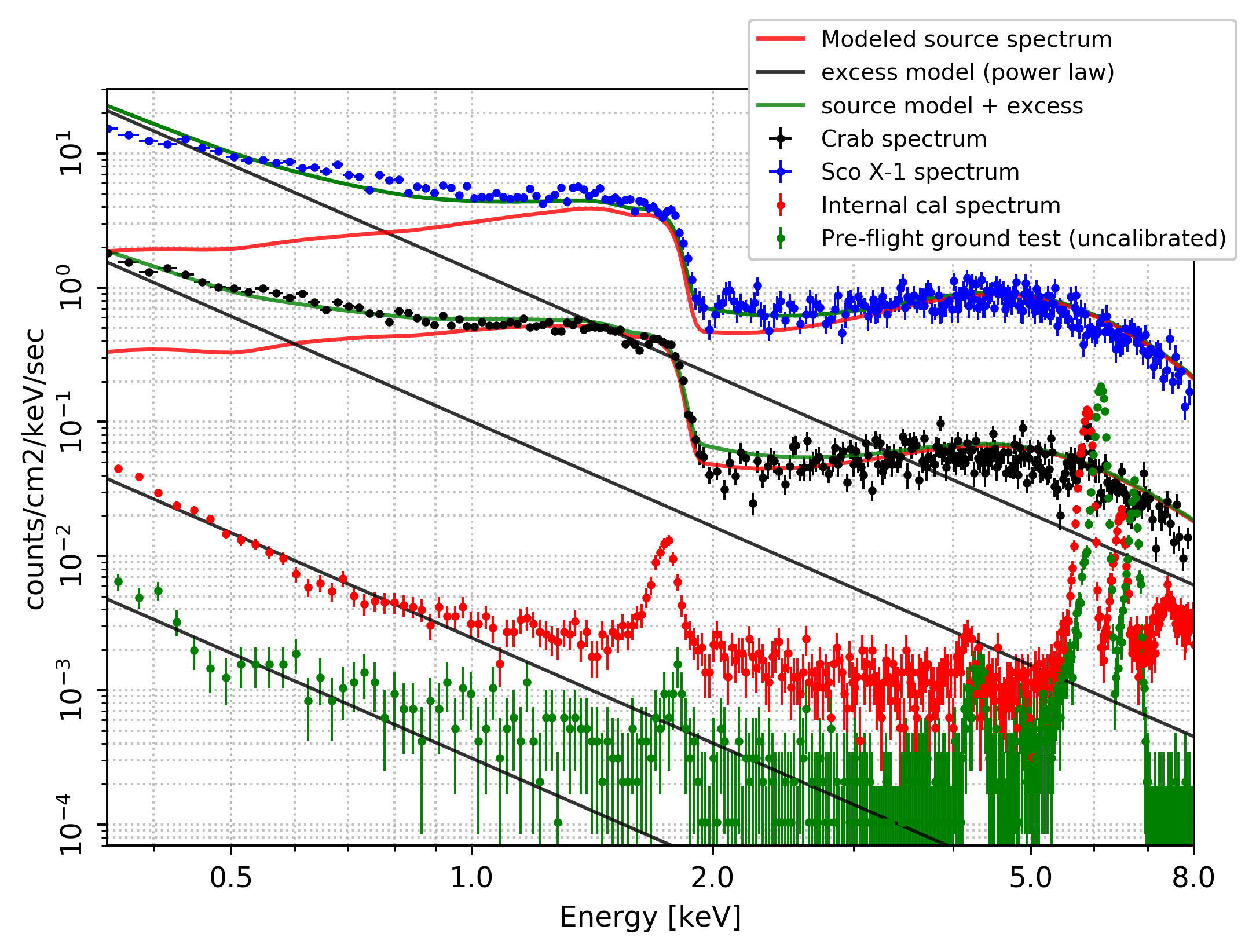}
	\caption{REXIS spectra from CCD2 node 2 illustrating an excess in X-ray events at low energies with a power law spectral shape. From top to bottom: observed spectra of Sco X-1 (blue dots), the Crab Nebula (black dots), Internal Calibration data (red dots), and uncalibrated data from ground tests prior to launch (green dots). For the Sco X-1 and Crab Nebula spectra, the expected spectra using the modeled instrument response are shown in the red solid lines, and models with an additional power law component are show in the green solid lines. The power law component (black solid line, index -2.6) is fit to the Crab data below 1~keV; for the Sco X-1, Internal Calibration, and ground test data, the power law  amplitude is scaled by the event rate observed in the 3-4~keV energy band. These results demonstrate that the excess at low energies scales directly with the incident X-ray flux.}
	\label{fig:spectra_LEE}
\end{figure}

One unexpected outcome of the Crab Calibration operation was the high count rates at energies below 1~keV, as shown in Figure~\ref{fig:ccd20_crab_fit}. The observed X-ray photon flux continues to increase as the energy decreases despite the fact that both the quantum efficiency of the REXIS CCDs and the X-ray flux of the Crab Nebula are known to decrease with lower energies. In fact, in order to explain the spectrum, the quantum efficiency of the REXIS CCDs would have to unphysically exceed 100\% at energies below 1 keV. The excess was observed again in the Mask Calibration operation, and was later confirmed to be present at much lower amplitude in the Internal Calibration data (collected before the radiation cover was opened) and in ground calibration data prior to launch; see Figure~\ref{fig:spectra_LEE}. 

Since the coded aperture mask shadow pattern results in only about the half of the pixels receiving photons illuminated from a point source, the low energy excess in the observed spectra of the Crab Nebula and Sco X-1 indicates that it originates from incident X-rays: i.e., the spatial information of the low-energy events correlates with the incident X-rays. This evidence, as well as results from ground tests and the observation that the magnitude of the excess scales with the incident X-ray flux, rules out several possibilities for noise, such as the cross-talk from CCD 3 (with the severed cable), the pipeline software or a previously unknown astrophysical effect. The excess pattern in the spectra is well-described by a power law model as seen in Figure~\ref{fig:spectra_LEE}. These features of the noise could be explained through the redistribution of true event energies to lower values, possibly due to charge loss in the CCD or an error in the flight software. A dead layer in the CCDs would introduce a similar effect, but models of the CCD response indicate that the thickness of the dead layer would have to be greater than 100~$\mu$m to explain the amount of the low energy excess, which contradicts the Q.E. at energies above 1~keV (where the expected CCD response accurately models the observed data). We observed similar low energy excess in the data taken with an Engineering Model (EM) CCD without an OBF, which rules out any issue with the OBF itself \citep[see also Figure 7 in][]{Koyama07}. Compton scattering of hard X-rays in the REXIS CCDs could generate low-energy events that are spatially correlated with X-rays from the Crab nebula or Sco X-1, but this contribution is not large enough to explain the magnitude of the excess. As of this writing, the REXIS instrument team does not have a complete explanation for the cause of the observed low energy excess. We continue to perform lab experiments using the flight spare and EM CCDs to find out the root cause of the low energy excess. Until the mechanism of the excess is understood and properly modeled, we have limited our analysis of REXIS data to energies above 1 keV.

\section{In-flight Calibration of XIS Imaging System} \label{sec:imaging}

The coded-aperture imaging technique used in REXIS is effectively a shadow-gram imaging system where the detector plane image is cross-correlated with the known mask pattern to identify the source position or to reconstruct the source shape. The angular resolution of REXIS is 26.5$'$, which is set by the telescope geometry (see Section~\ref{s:ang_resolution} in the Appendix). The localization error ($\delta\theta$) of a point source REXIS can achieve ranges from $\sim$ 1$'$ to $\sim$ 5$'$, with the instrument requirement being $\delta\theta < 6.7'$.

\subsection{Boresight Calibration through Observations of the Crab Nebula} \label{sec:boresight}

In order to resolve any regions of variable X-ray fluorescence across the surface of Bennu, we determined the precise position and orientation of the coded aperture mask relative to the REXIS CCDs using the observations of the Crab Nebula. For each CCD, we modeled the mask placement using four boresight parameters consisting of the three-dimensional positions and rotation (along the pointing axis) of the coded aperture mask relative to the CCD. In principle, two additional parameters describing the tilt of the mask are needed to fully describe the mask placement relative to the detector, but these parameters were not necessary in achieving the desired boresight corrections, as our measured accuracy was well within specifications indicating that there was no noticeable tilt in the mask.

As seen in Figure~\ref{f:crab_pointings}, the Crab Nebula observation sequence consisted of eight pointings surrounding the Crab Nebula.  These provided sufficient independent measurements to break the degeneracy in the boresight parameters while distributing flux evenly across the full detector area for absolute quantum efficiency calibration for all pixels. The detector plane image of each pointing was cross-correlated with the mask pattern through a series of Fast Fourier Transformation (FFTs) to reconstruct the matching sky image. Then, the bright point source in the image was localized through a Gaussian peak finding routine, and the offset from the true position of the Crab Nebula calculated. We repeated this process for a range of the values for each boresight parameter to solve for the correct value. The final boresight solution was obtained using a Markov Chain Monte Carlo (MCMC) search technique across the parameter space.

%For efficient optimization of the boresight parameters, we employed Markov Chain Monte Carlo (MCMC) simulations. First, we started the initial calculation with a set of  random values for each parameter within a reasonable range. Next, we selected a subset of the parameter values yielding the smallest offset errors, and then added a small perturbation to each value to get a new parameter value. In addition, we added a new set of  random values only within the range of the selected subset. We repeated the calculation with this new set of the parameters values. This approach converged to the our final boresight parameter values within several iterations.

\begin{SCfigure}
	\small
		\includegraphics[width=3in]{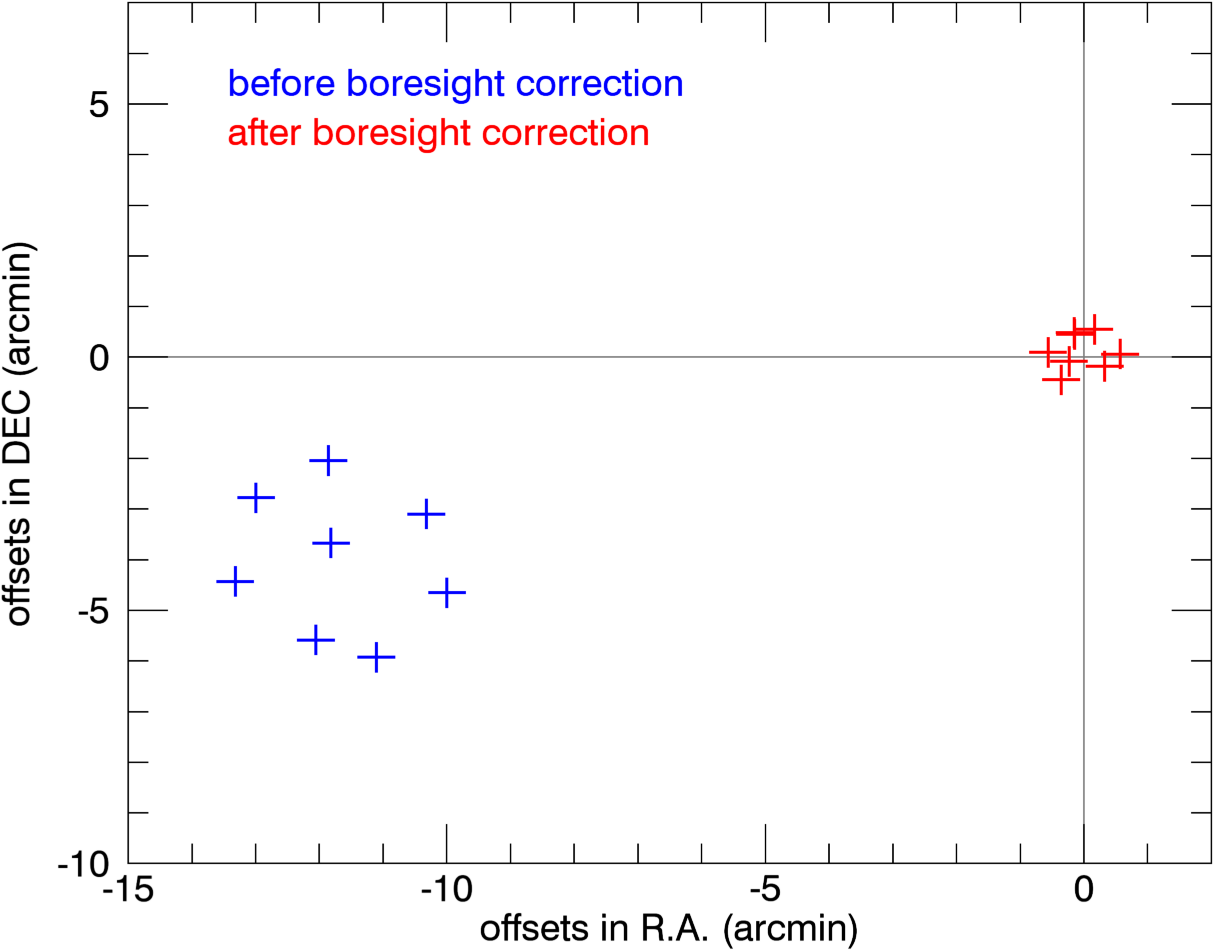}
	\caption{Offset between the measured and true positions of the Crab Nebula before (blue) and after (red) boresight corrections.}
	\label{f:boresight_errors}
\end{SCfigure}

Table~\ref{t:boresight_corrections} summarizes the boresight parameters using the first two days of the Crab Nebula observations in the 2nd Crab Calibration Operation. The measured values validated the REXIS instrument design and build indeed survived the launch and space environment within the original requirements. Figure~\ref{f:boresight_errors} shows the boresight errors before and after applying the boresight corrections using the parameter values in Table~\ref{t:boresight_corrections}. Before the correction, the raw pointing errors range $\sim$10$'$ to $\sim$15$'$ and after correction, the offsets are within 1$'$, which far exceeds the requirement.
Note that the signal-to-noise-ratio for each Crab Nebula observation is $>$20.

\begin{SCfigure}
	\small
		\includegraphics[width=3in]{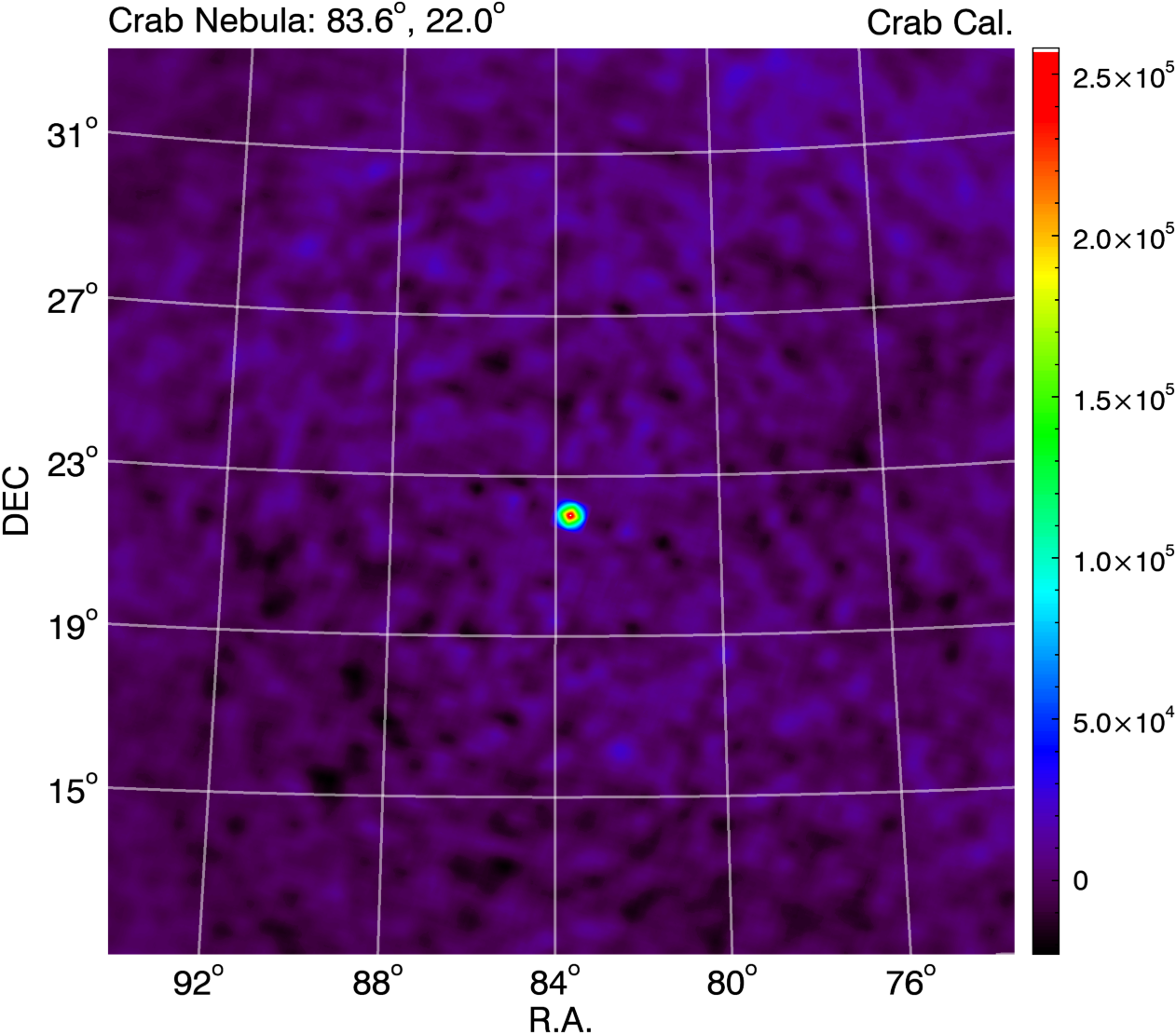}
	\caption{Reconstructed sky image from the first two day observations of the 2nd Crab Nebula calibration in March 2019. The color scale represent the total photon counts.}
	\label{f:crab_image}
\end{SCfigure}

\subsection{Mask Calibration through Observations of  Sco X-1} \label{sec:mask}

The REXIS boresight parameters were further confirmed with the observations of Sco X-1 obtained in June 2019. We observed Sco X-1 with five separate pointings (Figure~\ref{f:crab_pointings}) to see if there is any distortion in the mask after an operation that exposed the mask to direct sunlight. Direct exposure to sunlight potentially created a large temperature gradient within the mask, which could cause the mask to bow or stretch.  Unlike the pointings during the Crab Nebula observations, which were limited to 3.5$^\circ$ within the source, the five separate pointings during the Mask calibration include relatively large off-axis observations of the source, ranging up to $\sim$ 12$^\circ$ (Table~\ref{table:crab_observations}). 
Having the source at large off-axis angles enabled the inspection of outer mask segments. With the boresight corrections in Table~\ref{t:boresight_corrections},
the source positions were recovered within an arcminute of the true position. On the other hand, the data of two largest off pointings (2nd and 5th pointings of Sco X-1 in Table~\ref{table:crab_observations}) revealed a minor shrinkage in the observed mask pitch pixel in order of $\sim$1.5\%, which is equivalent to $\sim$ 25$''$. The observed shrinkage is well within the requirements for the mask and can be explained by a slightly bowing in the mask plane. However, it is not clear when the bowing was introduced since this was the first measurements for the mask distortion.

\begin{SCtable}
\small
\caption{Measured mask offsets relative to the REXIS CCDs vs.~the requirements (See Figure~\ref{f:instrument})}
\begin{tabular}{ c|c|c }
\hline\hline
Parameters & Measured & Requirements \\
\hline
$\Delta$X & $-$0.720 mm or $-$12.44$'$ & N/A  \\
$\Delta$Y & +0.023 mm  or +0.39$'$   & N/A  \\
$\Delta$Z & $-$0.891 mm              & $\pm$1.00 mm  \\
roll      & 0.205 degrees          &  0.44 degrees  \\
\hline
\end{tabular}
\label{t:boresight_corrections}
\end{SCtable}

\subsection{REXIS Detection of MAXI J0637-430} \label{s:maxi}

During the Orbital B observations, the new transient source MAXI J0637-430 was detected by the REXIS \citep{Allen20}. MAXI J0637-430 is a soft X-ray transient, first discovered on 2019 November 2 by the Monitor of All-sky X-ray Image (\MAXI), the Japanese Experiment on the International Space Station \citep{Negoro19}. The X-ray spectral and timing properties measured by the Nuclear Spectroscopic Telescope Array (\NuSTAR) along with the high X-ray to optical flux ratio suggests that the source is a black hole X-ray binary \citep{Tomsick19}. REXIS' detection of MAXI J0637-430 marks the first black hole detection made from outside the Earth-Moon environment.

\begin{SCfigure}
	\small
	    \includegraphics[width=3in]{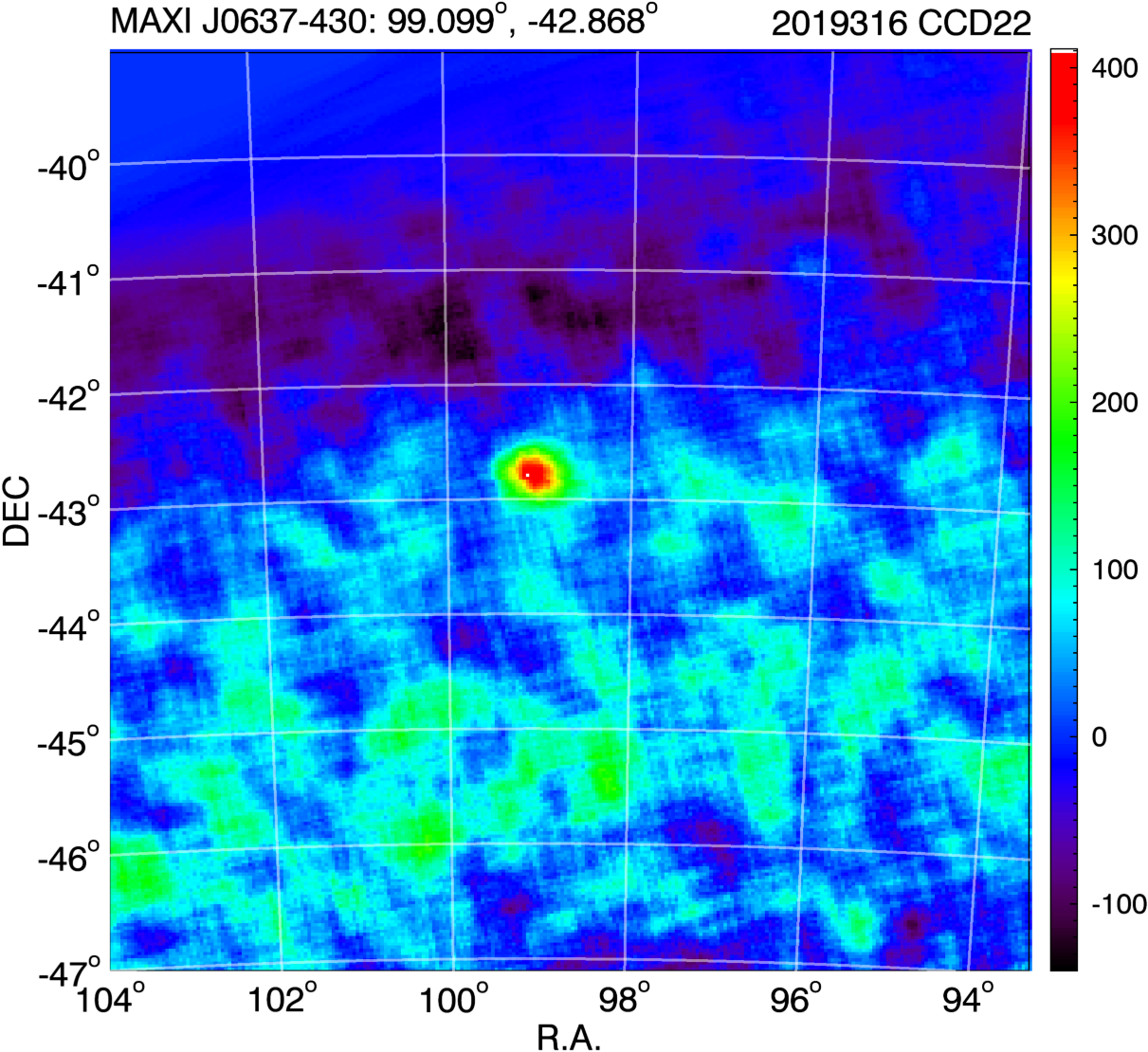}   
	\caption{Reconstructed REXIS sky image around the transient source MAX J0637-430 during Orbital R. The bright source seen in the image is 1.3$'$ away from the reported radio position of MAXI J0637-430.}
	\label{f:maxi_image}
\end{SCfigure}

The REXIS XIS detected the source (mainly by CCD 2) near the edge of the FoV just outside of Bennu for eight separate days between November 12 and 22 in 2019 (whenever it was in the FoV). 
With the boresight corrections in Table~\ref{t:boresight_corrections}, the REXIS position of the source is localized to (R.A., DEC) = (6:36:22.94, 	-42:50:44.0) (J2000), which is 1.3$'$ off the radio counterpart of the source \citep{Russell19}, based on the CCD 2 Node 2 data taken on November 12th, 2019. Figure~\ref{f:maxi_image} shows the reconstructed sky image from the data in CCD 2 Node 2, where the source dominantly shined.

%\FIX[in progress; analysis of the spectrum Cas A.]

\section{SXM Calibration}\label{sec:sxm}

As described in \citet{Masterson18}, the REXIS Solar X-ray Monitor (SXM) consists of a single Amptek XR-100 silicon drift diode (SDD). Due to a number of issues in the flight electronics, the energy resolution of the SXM was limited to 500~eV at 5.9~keV, and the low energy threshold was at approximately 1~keV under nominal operating conditions. Due to thermal sensitivity in the SXM electronics, the low energy threshold was approximately 2~keV during science operations around Bennu. The SXM field of view was defined by a 1~mm diameter collimator mounted above the SDD. The full width zero intensity (FWZI) field of view was 60$^\circ$, around the SXM boresight.  The SXM field of view was verified during solar pointing calibrations carried out before and after the OSIRIS-REx Earth Gravity Assist (EGA) in September 2017.

\begin{figure}
    \centering
    \includegraphics[width=0.99\textwidth]{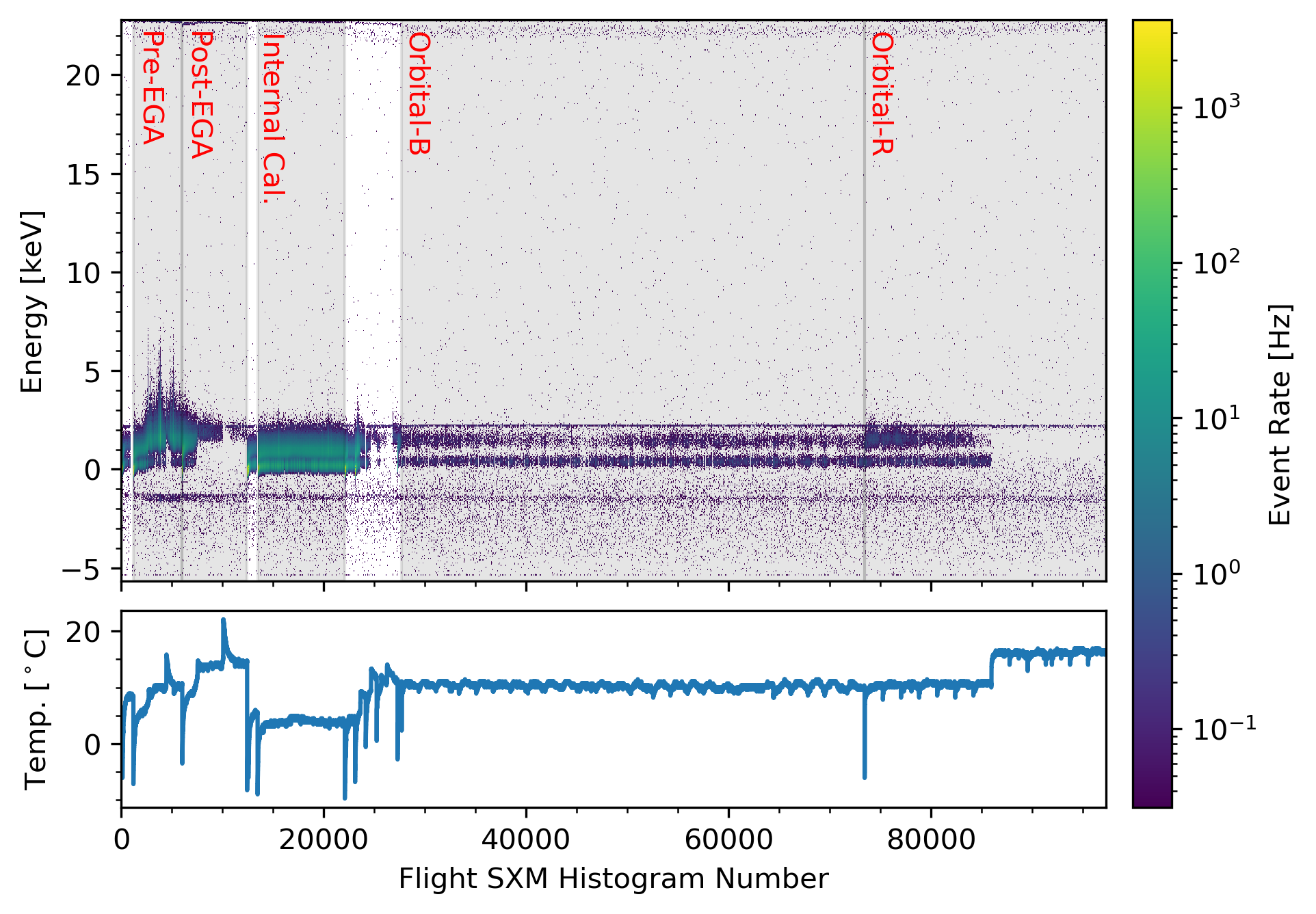}
    \caption{The time series of solar X-ray flux measured by the SXM throughout all REXIS operations is shown in the top panel, with temperature measurements shown below.  Increases in temperature in the MEB were observed to increase the low energy threshold of the detector which is most clearly visible in the pre- and post-EGA data (histogram number 4000-5000 and 7000-11000).  The "reset" artifacts used to monitor the stability of the SXM energy measurement are visible at approximately 75 ADU (-1.6 keV), 135 ADU (2.2 keV) during quiet solar states and near 500 ADU (22 keV) (see Figure \ref{fig:sxm_spectra}).  The intensity and position of these artificial lines enabled monitoring of the SXM gain throughout the flight.}
    \label{fig:sxm_temp_effects}
\end{figure}

\begin{figure}
    \centering
    \includegraphics[width=0.7\textwidth]{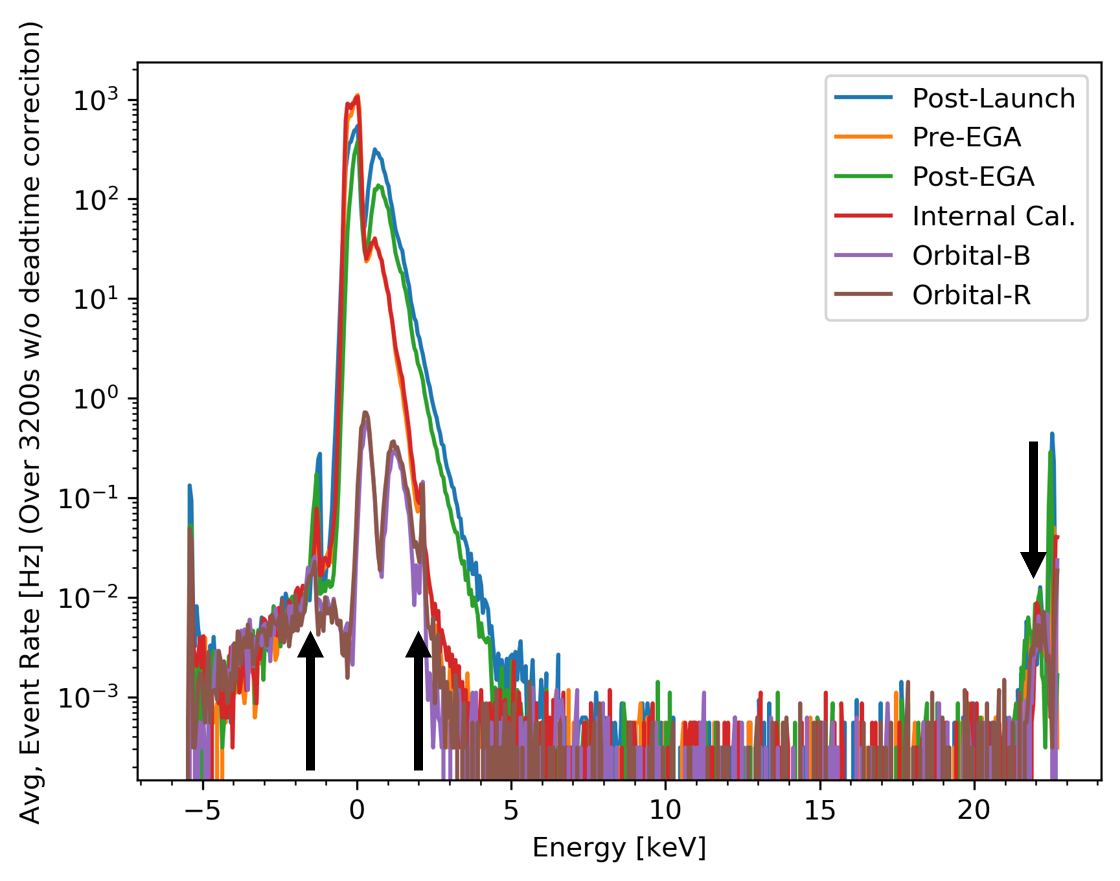}
    \caption{Individual SXM measurements of the solar spectrum at different times during the mission. The black arrows indicate the low energy and high energy reset artifacts used to monitor the gain of the SXM. Note the difference in the spectra acquired during Orbital B and R, compared to earlier operations. The solar X-ray flux was very quiet during Orbital B and R, and the high thermal load experienced by the MEB suppressed nearly all of the solar signal below 2~keV. During Orbital B and R thermal effects induced an increase in the energy threshold of the SXM leaving the data dominated by instrumental artifacts at 2 keV and below.}
    \label{fig:sxm_spectra}
\end{figure}

Since the SXM was not equipped with an on-board radioactive source for calibration, the gain stability of the instrument was monitored through occasional solar observations and by tracking electronics artifacts in the SXM spectrum (see Figures~\ref{fig:sxm_temp_effects} and \ref{fig:sxm_spectra}). These electronic ``reset" artifacts generated false events at fixed energies, and those fixed energies provided reference levels that allowed tracking of the gain. Changes in the energy resolution were tracked by measuring the high voltage bias, which remained nearly constant with less than 1\% change during the duration of the mission.

%Post-launch calibration of the SXM calibration was carried out during solar observations conducted prior to and following the OSIRIS-REx earth gravity assist (EGA) (hereafter pre- and post-EGA observations).  This enabled contemporaneous solar observations of the solar surface and corona from a viewing angle nearly identical to that observed from the earth.  This enabled cross-calibration of the SXM with Earth-orbiting assets in the extreme ultraviolet (EUV) and X-ray energies.   

The pre- and post-EGA observations enabled cross-calibration of the SXM with Earth-orbiting solar observatories, for example the Geostationary Operational Environmental Satellite (GOES)-16 Extreme Ultraviolet and X-ray Irradiance Sensors (EXIS) \citep{2009SPIE.7438E..02C} and the GOES-15 X-ray Sensor (XRS) \citep{1996SPIE.2812..344H}. The EXIS and XRS instruments observe the solar X-ray flux between 1.55 keV and 24 keV in two separate energy bands, denoted as A (1.55~--~12.4~keV) and B (3~--~24~keV). During the pre-EGA observation, a series of solar flares observed by the GOES instrument and the REXIS SXM enabled absolute calibration of the SXM effective area as noted in Table~\ref{t:intrument}.

Prior to launch, the REXIS team discovered that the SXM shaping electronics circuit on the Main Electronics Board (MEB) exhibited a thermal sensitivity that could increase the low energy threshold of the SXM.  That thermal sensitivity could change the threshold from the nominal setting of 0.3~keV up to values of approximately 2.5~keV. The effect was observed in the pre-EGA cross-calibration, when the SXM response showed intervals of non-correlated variability compared to data from the GOES instruments (see Figure \ref{fig:sxm_temp_effects}). The SXM had originally been designed to cover the energy range from 0.3 - 20~keV in order to characterize the incident solar spectrum above and below the nominal energy range of the REXIS CCDs.  However, due to the increased operating temperature during asteroid observations, the minimum operating energy threshold was shifted into the range of  1.8 -- 2 keV. The resulting loss of low energy sensitivity, coupled with low solar activity, severely limited the use of the SXM for the interpretation of data collected at Bennu.

%During the pre-EGA cross-calibration, the SXM response showed intervals of non-correlated variability compared to data from the GOES instruments (see figure \ref{fig:sxm_temp_effects}). The issue was ultimately traced to thermal effects within the shaping electronics circuit located on the Main Electronics Board (MEB).

%those reported by the GOES instrument at a few observation points during the pre-EGA calibration and during the majority of the post-EGA calibration.  The issue was traced back to thermal effects within the shaping electronics circuit located on the Main Electronics Board (MEB) housed at the base within the main REXIS spectrometer unit (see figure \ref{fig:sxm_temp_effects}).  Elevated temperatures within the main electronics box were observed to induce an increase in the low energy threshold of the SXM from the nominal setting of 1 keV up to approximately 2.5 keV.  This thermal sensitivity was known prior to launch and discovered just prior to delivery of the instrument in pre-delivery instrument-level thermal vacuum testing (TVAC).  Further oven-testing of the SXM instrument following TVAC thermal testing was carried out separately on the SXM SDD assembly (located near the high-gain antenna) while holding the MEB at room temperature enabling localization of the issue to the thermal condition of the MEB rather than being variable performance of solar pointed SDD sensor (see Allen et al. for details).  

\section{Summary and Lessons}\label{sec:summary}

REXIS was the first planetary X-ray instrument employing the coded-aperture imaging technique in hopes of spatially resolving variations in the elemental composition (if present) across the surface of an asteroid. It was an ambitious project for a student-led experiment involving a completely new planetary instrument design, necessarily conducted under a tight schedule and resource constraints. Spanning 9 years, from concept to completion of operations, more than 90 students had hands-on experience on the REXIS instrument development and operations, which also resulted in 12 Master and 2 Ph.~D. theses \citep{Masterson18}. 

Despite the naturally occurring challenge within a university of continual personnel changes from student rollovers as well as multiple challenges and delays during the instrument assembly, the in-flight operations and calibrations indicate that REXIS main spectrometer and solar X-ray monitor were built within their specifications and requirements, albeit with the lower effective detector area and a narrower energy range coverage than the original design. In particular, the observations of the Crab Nebula and Sco X-1 as well as the detection of the new black hole transient MAXI J0637-430 demonstrate that the REXIS imaging system was able to operate well within the original requirements. The historically low solar X-ray flux during Bennu science operations precluded detection of surface fluoresced X-ray photons \citep{Hoak21}.

% TRL description

The successful operation of REXIS during calibration and the detection of the MAXI source while observing Bennu enables the extension of the Technology Readiness Level (TRL) to interplanetary operation for several key subsystems in REXIS. First, the REXIS CCDs are the same type of CCDs (MIT/LL CCID-41) flown on the JAXA {\it Suzaku} mission, which have been operating in low-Earth orbit for about 15 years. REXIS extends the TRL of the MIT CCID-41 CCDs to the interplanetary and near-asteroid environment. Second, the successful detection of celestial X-ray sources such as MAXI J0637-430 under bright background light from Bennu boosted the maturity of the directly deposited OBF in the REXIS CCDs to TRL 9 \citep[see also][]{Thayer21}. Third, the successful opening of the radiation cover in the first attempt after a 2-year traverse across interplanetary space shows the one-time release TiNi Aerospace FD04 Frangibolt actuator at TRL 9 for use after a long duration cruise.  Prior to REXIS, this model Frangibolt had been, to our knowledge, flown only in low-Earth  orbit \citep{Masterson18}. Fourth, the detector in the REXIS SXM Amptek XR-100 SDD) is the same type used in several space missions such as \NICER \citep{Gendreau16} and the Miniature X-ray Solar Spectrometer (MinXSS) \citep{Woods17}. Despite issues in the readout system, the REXIS SDD functioned properly for the first two years, extending the TRL of the Amptek SDD to the interplanetary environment.
%, similar to the KETEK VITUS SDDs mounted on Mars rover “Perseverance”.

REXIS suffered from two common causes of performance degradation in space instruments: inadequate light leak protection and thermal management challenges. Although the REXIS team was acutely aware of these potential issues, limits on pre-flight testing and calibration due to schedule constraints prevented the team from addressing certain vulnerabilities that could have been more fully recognized and addressed before launch. In-flight calibrations of the REXIS main spectrometer revealed that the CCDs in REXIS were susceptible to light leaks from their sides. We note that in contrast the optical blocking filter (OBF) on the incident surface of the CCDs performed excellently. During the REXIS observations of Bennu, these light leaks further reduced the effective integration time for the science objective of detecting X-rays emitted from the surface of Bennu.  With these challenges to the instrument itself, the unexpectedly low solar X-ray emission (even considering solar minimum) during the Orbital B and Orbital R observations  prevented the detection of X-ray fluorescence from Bennu's surface \citep{Hoak21}.
In the case of the SXM, it performed as expected until the Orbital B observation period, at which time it lost its low energy response due to the unexpected out-of-bound thermal environment experienced by the main electronics board. 

%The late discovery of the issue in the SXM readout electronics severely limited the range of mitigations that the team could take.

%Some of the issues that contributed to non-detection of the XRF signal from Bennu could have been mitigated 
%from the pre-flight testing and calibrations. In the case of REXIS, the tight schedule and the assembly delay prevented most of the pre-planned calibration and testing for REXIS.

Here we list some of key pre-flight activities that could have improved REXIS performance, offered as lessons for future planetary X-ray spectroscopy investigations. Given the delays during the instrument development and assembly, 
it would have been challenging to conduct these activities prior launch even with the current knowledge of REXIS performance. However, we recommend a careful consideration of the priority of these activities in the context of other pre-flight tasks for future missions.

First, more in-depth tests (and modeling) of light leaks to ensure low background noise for the CCDs. REXIS employed a new type of OBF for the first time for X-ray CCDs in space, where an Al layer is directly deposited on the top surface of the CCDs. As a result, the pre-flight activities regarding light leaks focused on the OBF design and performance. The preliminary light leak modeling indeed revealed possible light-leak paths from the side and back of the CCDs. This discovery led to a few design changes, but these proved insufficient. Given the complexity involved in modeling pathways for light leaks in general, the light leak modeling can be prohibitively expensive and often ineffective in reflecting the reality. Instead, the lab experience using a separate engineering model (EM) setup tailored for light leak tests could be an efficient way to improve the fidelity in the light leak modeling and to validate any design changes if needed.

Second,  one needs a full characterization of the CCD responses at low energies to set the key operational parameters such as event and split charge thresholds. Without the full characterization of the instrument response, the REXIS team relied on in-flight operations for both calibrations and parameter optimizations. This not only led to a limited success in the parameter optimization (e.g., the adjustment for the split charge threshold was made after the Orbital B observations) but also added a burden for the \OSIRIS operational team to accommodate various tests for the REXIS system.  While those accommodations were made graciously and with enthusiasm for support of the REXIS students, they were an added weight nonetheless.

Third, we would have benefited from a more thorough thermal modeling and testing of the REXIS system to ensure successful operation for a wider range of temperatures.
The late discovery of the issue in the SXM readout electronics severely limited the range of mitigations that the team could take. The adopted solution still left the system with the limited range of the threshold setting, but more importantly the associated delay severely limited pre-flight testing of the SXM system. Relying on thermal tests of the full or near-complete system is not ideal in terms of schedule. Repeating the test can be also taxing in  terms of both the schedule and resources. One could consider a piece-wise or independent thermal test of the MEB or other key components at a relatively early phase with the help of the appropriate diagnostic features in the boards.

Fourth, REXIS was limited by a deficit of in-depth testing and debugging of the flight software, along with originally planned but never-implemented software features. An example is the option to downlink a full 3 $\times$ 3 pixel data array for each CCD event. In-flight grading for CCD events was adopted for REXIS to minimize the telemetry rate, but the accuracy of in-flight grading can vary with spatial and temporal response changes in the CCDs. Inaccurate onboard grading can result in degradation in the spectral resolution and quantum efficiency, since the recovery of the precise energy of input X-rays for multi-pixel events would require all the split signals to be properly selected and calibrated. The downlink of the full 3 $\times$ 3 pixel energies for each event could enable further improvement in energy resolution and quantum efficiency of single pixel events from ground analysis (e.g., the split charge threshold can be readjusted during the analysis).

Finally, a dedicated calibration operation or calibration source for low energy lines 
would have been also helpful. Besides MAXI J0637-430, REXIS has detected a few other X-ray sources during the Orbital B and R observations.  Among these, Cassiopeia A (Cas A)
Supernova Remnant is interesting due to its prominent line features at low energies in the X-ray spectrum \cite[e.g.,][]{1997AA...324L..49F}.  
The reconstructed sky image shows a clear source 
at the expected position of Cas A. The measured background subtracted REXIS spectrum show only marginal line features, 
while the overall continuum flux level matches with the expected values. During the Orbital B and R observations, 
these celestial sources including Cas A were detected in the peripheral of the REXIS FoV with relatively limited
exposure, and the low photon statistics indicates that lack of prominent line features in the measured spectrum is in fact consistent with what is expected.

\begin{acknowledgements}
We thank the \OSIRIS team for accommodating the REXIS team's various operational requests as well as their helpful suggestions and guidance.
The authors would like to thank Beth Clark and the members of the \OSIRIS X-ray Science Working Group for valuable advice and support.  
The REXIS instrument was built by a student team that included over eighty graduate and undergraduate students with invaluable support from Mark Bautz, Joel Villasenor and Steve Kissel from MIT Kavli Institute, the \OSIRIS payload team at NASA GFSC, especially Mary Walker, Libby Adelman, Michael Choi, James Dailey, Michael Pryzby and David Petrick, Lockheed Martin payload engineer Jill Cattrysse-Larson, and members of the \OSIRIS Standing Review Board, Steven Battel, Ed Powers and Mark Kahan. 
We also thank Dr.~Keith Gendreau and the \NICER team for enabling the near-simultaneous observations of Sco X-1 with \NICER and
Dr. Jeroen Homan for assisting the analysis of the \NICER data.
This work is supported by NASA Grant NNM10AA11C.
MIT Lincoln Laboratory material is based upon work supported by the United States Air Force under Air Force Contract No. FA8702-15-D-0001. Any opinions, findings, conclusions or recommendations expressed in this material are those of the author(s) and do not necessarily reflect the views of the United States Air Force. This document is approved for public release. Distribution is unlimited.

\end{acknowledgements}

\section*{Conflict of interest}
The authors declare that they have no conflict of interest.

\bibliographystyle{spbasic}      % basic style, author-year citations

\bibliography{references.bib}

\begin{thebibliography}{28}
\providecommand{\natexlab}[1]{#1}
\providecommand{\url}[1]{{#1}}
\providecommand{\urlprefix}{URL }
\expandafter\ifx\csname urlstyle\endcsname\relax
  \providecommand{\doi}[1]{DOI~\discretionary{}{}{}#1}\else
  \providecommand{\doi}{DOI~\discretionary{}{}{}\begingroup
  \urlstyle{rm}\Url}\fi
\providecommand{\eprint}[2][]{\url{#2}}

\bibitem[{{Allen} et~al.(2013){Allen}, {Grindlay}, {Hong}, {Binzel},
  {Masterson}, {Inamdar}, {Chodas}, {Smith}, {Bautz}, {Kissel}, {Villasenor},
  {Oprescu}, and {Induni}}]{Allen13}
{Allen} B, {Grindlay} J, {Hong} J, {Binzel} RP, {Masterson} R, {Inamdar} NK,
  {Chodas} M, {Smith} MW, {Bautz} MW, {Kissel} SE, {Villasenor} J, {Oprescu} M,
  {Induni} N (2013) {The REgolith X-Ray Imaging Spectrometer (REXIS) for
  OSIRIS-REx: identifying regional elemental enrichment on asteroids}. In:
  {Kahan} MA, {Levine} MB (eds) Optical Modeling and Performance Predictions
  VI, Society of Photo-Optical Instrumentation Engineers (SPIE) Conference
  Series, vol 8840, p 88400M, \doi{10.1117/12.2041715}, \eprint{1309.6665}

\bibitem[{{Allen} et~al.(2020){Allen}, {Grindlay}, {Hoak}, {Hong}, {Guevel},
  {Lambert}, {Binzel}, {Masterson}, {Cummings}, {Lim}, {Lauretta}, and
  {Boynton}}]{Allen20}
{Allen} B, {Grindlay} J, {Hoak} D, {Hong} J, {Guevel} D, {Lambert} M, {Binzel}
  RP, {Masterson} R, {Cummings} A, {Lim} L, {Lauretta} D, {Boynton} B (2020)
  {Detection of MAXI J0637-430 by the Regolith X-Ray Imaging Spectrometer
  (REXIS) Onboard OSIRIS-REx}. The Astronomer's Telegram 13594:1

\bibitem[{{Bautz} et~al.(2004){Bautz}, {Kissel}, {Prigozhin}, {LaMarr},
  {Burke}, and {Gregory}}]{Bautz04}
{Bautz} MW, {Kissel} SE, {Prigozhin} GY, {LaMarr} B, {Burke} BE, {Gregory} JA
  (2004) {Progress in x-ray CCD sensor performance for the Astro-E2 X-ray
  imaging spectrometer}. In: {Holland} AD (ed) High-Energy Detectors in
  Astronomy, Society of Photo-Optical Instrumentation Engineers (SPIE)
  Conference Series, vol 5501, pp 111--122, \doi{10.1117/12.553198}

\bibitem[{{Biswas}(2016)}]{Biswas16}
{Biswas} P (2016) {Radiation management, avionics development, and integrated
  testing of a class-D space-based asteroid X-ray spectrometer}. Masters
  Thesis, Massachusetts Institute of Technology, Cambridge

\bibitem[{{Chamberlin} et~al.(2009){Chamberlin}, {Woods}, {Eparvier}, and
  {Jones}}]{2009SPIE.7438E..02C}
{Chamberlin} PC, {Woods} TN, {Eparvier} FG, {Jones} AR (2009) {Next generation
  x-ray sensor (XRS) for the NOAA GOES-R satellite series}. In: {Fineschi} S,
  {Fennelly} JA (eds) Solar Physics and Space Weather Instrumentation III,
  Society of Photo-Optical Instrumentation Engineers (SPIE) Conference Series,
  vol 7438, p 743802, \doi{10.1117/12.826807}

\bibitem[{{Dicke}(1968)}]{Dicke68}
{Dicke} RH (1968) {Scatter-Hole Cameras for X-Rays and Gamma Rays}. The
  Astrophysical Journal Letters 153:L101, \doi{10.1086/180230}

\bibitem[{{Favata} et~al.(1997){Favata}, {Vink}, {dal Fiume}, {Parmar},
  {Santangelo}, {Mineo}, {Preite-Martinez}, {Kaastra}, and
  {Bleeker}}]{1997AA...324L..49F}
{Favata} F, {Vink} J, {dal Fiume} D, {Parmar} AN, {Santangelo} A, {Mineo} T,
  {Preite-Martinez} A, {Kaastra} JS, {Bleeker} JAM (1997) {The broad-band X-ray
  spectrum of the CAS A supernova remnant as seen by the BeppoSAX observatory.}
  Astronomy and Astrophysics 324:L49--L52, \eprint{astro-ph/9707052}

\bibitem[{{Freeman} et~al.(2001){Freeman}, {Doe}, and
  {Siemiginowska}}]{Freeman01}
{Freeman} P, {Doe} S, {Siemiginowska} A (2001) {Sherpa: a mission-independent
  data analysis application}. Proc SPIE 4477(76), \doi{10.1117/12.447161}

\bibitem[{{Gendreau} et~al.(2016){Gendreau}, {Arzoumanian}, {Adkins}, {Albert},
  {Anders}, {Aylward}, {Baker}, {Balsamo}, {Bamford}, {Benegalrao}, {Berry},
  {Bhalwani}, {Black}, {Blaurock}, {Bronke}, {Brown}, {Budinoff}, {Cantwell},
  {Cazeau}, {Chen}, {Clement}, {Colangelo}, {Coleman}, {Coopersmith},
  {Dehaven}, {Doty}, {Egan}, {Enoto}, {Fan}, {Ferro}, {Foster}, {Galassi},
  {Gallo}, {Green}, {Grosh}, {Ha}, {Hasouneh}, {Heefner}, {Hestnes}, {Hoge},
  {Jacobs}, {J{\o}rgensen}, {Kaiser}, {Kellogg}, {Kenyon}, {Koenecke}, {Kozon},
  {LaMarr}, {Lambertson}, {Larson}, {Lentine}, {Lewis}, {Lilly}, {Liu},
  {Malonis}, {Manthripragada}, {Markwardt}, {Matonak}, {Mcginnis}, {Miller},
  {Mitchell}, {Mitchell}, {Mohammed}, {Monroe}, {Montt de Garcia}, {Mul{\'e}},
  {Nagao}, {Ngo}, {Norris}, {Norwood}, {Novotka}, {Okajima}, {Olsen},
  {Onyeachu}, {Orosco}, {Peterson}, {Pevear}, {Pham}, {Pollard}, {Pope},
  {Powers}, {Powers}, {Price}, {Prigozhin}, {Ramirez}, {Reid}, {Remillard},
  {Rogstad}, {Rosecrans}, {Rowe}, {Sager}, {Sanders}, {Savadkin}, {Saylor},
  {Schaeffer}, {Schweiss}, {Semper}, {Serlemitsos}, {Shackelford}, {Soong},
  {Struebel}, {Vezie}, {Villasenor}, {Winternitz}, {Wofford}, {Wright}, {Yang},
  and {Yu}}]{Gendreau16}
{Gendreau} KC, {Arzoumanian} Z, {Adkins} PW, {Albert} CL, {Anders} JF,
  {Aylward} AT, {Baker} CL, {Balsamo} ER, {Bamford} WA, {Benegalrao} SS,
  {Berry} DL, {Bhalwani} S, {Black} JK, {Blaurock} C, {Bronke} GM, {Brown} GL,
  {Budinoff} JG, {Cantwell} JD, {Cazeau} T, {Chen} PT, {Clement} TG,
  {Colangelo} AT, {Coleman} JS, {Coopersmith} JD, {Dehaven} WE, {Doty} JP,
  {Egan} MD, {Enoto} T, {Fan} TW, {Ferro} DM, {Foster} R, {Galassi} NM, {Gallo}
  LD, {Green} CM, {Grosh} D, {Ha} KQ, {Hasouneh} MA, {Heefner} KB, {Hestnes} P,
  {Hoge} LJ, {Jacobs} TM, {J{\o}rgensen} JL, {Kaiser} MA, {Kellogg} JW,
  {Kenyon} SJ, {Koenecke} RG, {Kozon} RP, {LaMarr} B, {Lambertson} MD, {Larson}
  AM, {Lentine} S, {Lewis} JH, {Lilly} MG, {Liu} KA, {Malonis} A,
  {Manthripragada} SS, {Markwardt} CB, {Matonak} BD, {Mcginnis} IE, {Miller}
  RL, {Mitchell} AL, {Mitchell} JW, {Mohammed} JS, {Monroe} CA, {Montt de
  Garcia} KM, {Mul{\'e}} PD, {Nagao} LT, {Ngo} SN, {Norris} ED, {Norwood} DA,
  {Novotka} J, {Okajima} T, {Olsen} LG, {Onyeachu} CO, {Orosco} HY, {Peterson}
  JR, {Pevear} KN, {Pham} KK, {Pollard} SE, {Pope} JS, {Powers} DF, {Powers}
  CE, {Price} SR, {Prigozhin} GY, {Ramirez} JB, {Reid} WJ, {Remillard} RA,
  {Rogstad} EM, {Rosecrans} GP, {Rowe} JN, {Sager} JA, {Sanders} CA, {Savadkin}
  B, {Saylor} MR, {Schaeffer} AF, {Schweiss} NS, {Semper} SR, {Serlemitsos} PJ,
  {Shackelford} LV, {Soong} Y, {Struebel} J, {Vezie} ML, {Villasenor} JS,
  {Winternitz} LB, {Wofford} GI, {Wright} MR, {Yang} MY, {Yu} WH (2016) {The
  Neutron star Interior Composition Explorer (NICER): design and development}.
  In: {den Herder} JWA, {Takahashi} T, {Bautz} M (eds) Space Telescopes and
  Instrumentation 2016: Ultraviolet to Gamma Ray, Society of Photo-Optical
  Instrumentation Engineers (SPIE) Conference Series, vol 9905, p 99051H,
  \doi{10.1117/12.2231304}

\bibitem[{{Hanser} and {Sellers}(1996)}]{1996SPIE.2812..344H}
{Hanser} FA, {Sellers} FB (1996) {Design and calibration of the GOES-8 solar
  x-ray sensor: the XRS}. In: {Washwell} ER (ed) GOES-8 and Beyond, Society of
  Photo-Optical Instrumentation Engineers (SPIE) Conference Series, vol 2812,
  pp 344--352, \doi{10.1117/12.254082}

\bibitem[{{Hoak} et~al.(2021){Hoak}, {Binzel}, {Allen}, {Hong}, {Guevel},
  {Grindlay}, {Masterson}, {Chodas}, {Cummings}, {Lambert}, {Thayer}, {Lim},
  {Clark}, {McCoy}, and {Lauretta}}]{Hoak21}
{Hoak} D, {Binzel} RP, {Allen} B, {Hong} J, {Guevel} D, {Grindlay} J,
  {Masterson} R, {Chodas} M, {Cummings} A, {Lambert} M, {Thayer} C, {Lim} LF,
  {Clark} BE, {McCoy} TJ, {Lauretta} D (2021) {Results from the REgolith X-ray
  Imaging Spectrometer (REXIS) at Bennu}. In Preparation

\bibitem[{{Jones} et~al.(2014){Jones}, {Chodas}, {Smith}, and
  {Masterson}}]{Jones14}
{Jones} M, {Chodas} M, {Smith} MJ, {Masterson} RA (2014) {Engineering design of
  the Regolith X-ray Imaging Spectrometer (REXIS) instrument: an OSIRIS-REx
  student collaboration}. In: {Takahashi} T, {den Herder} JWA, {Bautz} M (eds)
  Space Telescopes and Instrumentation 2014: Ultraviolet to Gamma Ray, Society
  of Photo-Optical Instrumentation Engineers (SPIE) Conference Series, vol
  9144, p 914453, \doi{10.1117/12.2056903}

\bibitem[{{Koyama} et~al.(2007){Koyama}, {Tsunemi}, {Dotani}, {Bautz},
  {Hayashida}, {Tsuru}, {Matsumoto}, {Ogawara}, {Ricker}, {Doty}, {Kissel},
  {Foster}, {Nakajima}, {Yamaguchi}, {Mori}, {Sakano}, {Hamaguchi},
  {Nishiuchi}, {Miyata}, {Torii}, {Namiki}, {Katsuda}, {Matsuura}, {Miyauchi},
  {Anabuki}, {Tawa}, {Ozaki}, {Murakami}, {Maeda}, {Ichikawa}, {Prigozhin},
  {Boughan}, {Lamarr}, {Miller}, {Burke}, {Gregory}, {Pillsbury}, {Bamba},
  {Hiraga}, {Senda}, {Katayama}, {Kitamoto}, {Tsujimoto}, {Kohmura}, {Tsuboi},
  and {Awaki}}]{Koyama07}
{Koyama} K, {Tsunemi} H, {Dotani} T, {Bautz} MW, {Hayashida} K, {Tsuru} TG,
  {Matsumoto} H, {Ogawara} Y, {Ricker} GR, {Doty} J, {Kissel} SE, {Foster} R,
  {Nakajima} H, {Yamaguchi} H, {Mori} H, {Sakano} M, {Hamaguchi} K, {Nishiuchi}
  M, {Miyata} E, {Torii} K, {Namiki} M, {Katsuda} S, {Matsuura} D, {Miyauchi}
  T, {Anabuki} N, {Tawa} N, {Ozaki} M, {Murakami} H, {Maeda} Y, {Ichikawa} Y,
  {Prigozhin} GY, {Boughan} EA, {Lamarr} B, {Miller} ED, {Burke} BE, {Gregory}
  JA, {Pillsbury} A, {Bamba} A, {Hiraga} JS, {Senda} A, {Katayama} H,
  {Kitamoto} S, {Tsujimoto} M, {Kohmura} T, {Tsuboi} Y, {Awaki} H (2007) {X-Ray
  Imaging Spectrometer (XIS) on Board Suzaku}. Publications of the Astronomical
  Society of Japan 59:23--33, \doi{10.1093/pasj/59.sp1.S23}

\bibitem[{{Lambert}(2020)}]{Lambert20}
{Lambert} MM (2020) {A root cause analysis of REXIS detection efficiency loss
  during phase E operations}. Masters Thesis, Massachusetts Institute of
  Technology, Cambridge

\bibitem[{{Masterson} et~al.(2018){Masterson}, {Chodas}, {Bayley}, {Allen},
  {Hong}, {Biswas}, {McMenamin}, {Stout}, {Bokhour}, {Bralower}, {Carte},
  {Chen}, {Jones}, {Kissel}, {Schmidt}, {Smith}, {Sondecker}, {Lim},
  {Lauretta}, {Grindlay}, and {Binzel}}]{Masterson18}
{Masterson} RA, {Chodas} M, {Bayley} L, {Allen} B, {Hong} J, {Biswas} P,
  {McMenamin} C, {Stout} K, {Bokhour} E, {Bralower} H, {Carte} D, {Chen} S,
  {Jones} M, {Kissel} S, {Schmidt} F, {Smith} M, {Sondecker} G, {Lim} LF,
  {Lauretta} DS, {Grindlay} JE, {Binzel} RP (2018) {Regolith X-Ray Imaging
  Spectrometer (REXIS) Aboard the OSIRIS-REx Asteroid Sample Return Mission}.
  Space Science Reviews 214(1):48, \doi{10.1007/s11214-018-0483-8}

\bibitem[{{Negoro} et~al.(2019){Negoro}, {Miike}, {Nakajima}, {Maruyama},
  {Aoki}, {Kobayashi}, {Mihara}, {Tamagawa}, {Matsuoka}, {Sakamoto}, {Serino},
  {Sugita}, {Nishida}, {Yoshida}, {Tsuboi}, {Iwakiri}, {Sasaki}, {Kawai},
  {Sato}, {Shidatsu}, {Kawai}, {Oeda}, {Shiraishi}, {Nakahira}, {Sugawara},
  {Ueno}, {Tomida}, {Ishikawa}, {Isobe}, {Shimomukai}, {Tominaga}, {Ueda},
  {Tanimoto}, {Yamada}, {Ogawa}, {Setoguchi}, {Yoshitake}, {Tsunemi},
  {Yoneyama}, {Asakura}, {Hattori}, {Yamauchi}, {Iwahori}, {Kurihara},
  {Kurogi}, {Kawamuro}, {Yamaoka}, {Kawakubo}, {Sugizaki}, and {MAXI
  Team}}]{Negoro19}
{Negoro} H, {Miike} K, {Nakajima} M, {Maruyama} W, {Aoki} M, {Kobayashi} K,
  {Mihara} T, {Tamagawa} T, {Matsuoka} M, {Sakamoto} T, {Serino} M, {Sugita} S,
  {Nishida} H, {Yoshida} A, {Tsuboi} Y, {Iwakiri} W, {Sasaki} R, {Kawai} H,
  {Sato} T, {Shidatsu} M, {Kawai} N, {Oeda} M, {Shiraishi} K, {Nakahira} S,
  {Sugawara} Y, {Ueno} S, {Tomida} H, {Ishikawa} M, {Isobe} N, {Shimomukai} R,
  {Tominaga} M, {Ueda} Y, {Tanimoto} A, {Yamada} S, {Ogawa} S, {Setoguchi} K,
  {Yoshitake} T, {Tsunemi} H, {Yoneyama} T, {Asakura} K, {Hattori} K,
  {Yamauchi} M, {Iwahori} S, {Kurihara} Y, {Kurogi} K, {Kawamuro} T, {Yamaoka}
  K, {Kawakubo} Y, {Sugizaki} M, {MAXI Team} (2019) {MAXI/GSC detection of a
  soft X-ray transient MAXI J0637-430}. The Astronomer's Telegram 13256:1

\bibitem[{{Nittler} et~al.(2001){Nittler}, {Starr}, {Lim}, {McCoy}, {Burbine},
  {Reedy}, {Trombka}, {Gorenstein}, {Squyres}, {Boynton}, {McClanahan},
  {Bhangoo}, {Clark}, {Murphy}, and {Killen}}]{Nittler01}
{Nittler} LR, {Starr} RD, {Lim} L, {McCoy} TJ, {Burbine} TH, {Reedy} RC,
  {Trombka} JI, {Gorenstein} P, {Squyres} SW, {Boynton} WV, {McClanahan} TP,
  {Bhangoo} JS, {Clark} PE, {Murphy} ME, {Killen} R (2001) {X-ray fluorescence
  measurements of the surface elemental composition of asteroid 433 Eros}.
  Meteoritics and Planetary Science 36(12):1673--1695,
  \doi{10.1111/j.1945-5100.2001.tb01856.x}

\bibitem[{{Okada} et~al.(2006){Okada}, {Shirai}, {Yamamoto}, {Arai}, {Ogawa},
  {Hosono}, and {Kato}}]{Okada06}
{Okada} T, {Shirai} K, {Yamamoto} Y, {Arai} T, {Ogawa} K, {Hosono} K, {Kato} M
  (2006) {X-ray Fluorescence Spectrometry of Asteroid Itokawa by Hayabusa}.
  Science 312(5778):1338--1341, \doi{10.1126/science.1125731}

\bibitem[{{Prigozhin} et~al.(2000){Prigozhin}, {Jones}, {Bautz}, {Ricker}, and
  {Kraft}}]{Prigozhin00}
{Prigozhin} G, {Jones} S, {Bautz} M, {Ricker} G, {Kraft} S (2000) {The physics
  of the low-energy tail in the ACIS CCD. The spectral redistribution
  function}. Nuclear Instruments and Methods in Physics Research A
  439(2-3):582--591, \doi{10.1016/S0168-9002(99)00850-5}

\bibitem[{{Prigozhin} et~al.(1998){Prigozhin}, {Rasmussen}, {Bautz}, and
  {Ricker}}]{Prigozhin98}
{Prigozhin} GY, {Rasmussen} A, {Bautz} MW, {Ricker} GR (1998) {Model of the
  x-ray response of the ACIS CCD}. In: {Hoover} RB, {Walker} AB (eds) X-Ray
  Optics, Instruments, and Missions, Society of Photo-Optical Instrumentation
  Engineers (SPIE) Conference Series, vol 3444, pp 267--275,
  \doi{10.1117/12.331291}

\bibitem[{{Prigozhin} et~al.(2003){Prigozhin}, {Bautz}, and
  {Ricker}}]{Prigozhin03}
{Prigozhin} GY, {Bautz} MW, {Ricker} J George~R (2003) {Model of the backside
  illuminated Chandra CCD}. In: {Truemper} JE, {Tananbaum} HD (eds) X-Ray and
  Gamma-Ray Telescopes and Instruments for Astronomy., Society of Photo-Optical
  Instrumentation Engineers (SPIE) Conference Series, vol 4851, pp 149--156,
  \doi{10.1117/12.461428}

\bibitem[{{Rizk} et~al.(2018){Rizk}, {Drouet d'Aubigny}, {Golish}, {Fellows},
  {Merrill}, {Smith}, {Walker}, {Hendershot}, {Hancock}, {Bailey},
  {DellaGiustina}, {Lauretta}, {Tanner}, {Williams}, {Harshman}, {Fitzgibbon},
  {Verts}, {Chen}, {Connors}, {Hamara}, {Dowd}, {Lowman}, {Dubin}, {Burt},
  {Whiteley}, {Watson}, {McMahon}, {Ward}, {Booher}, {Read}, {Williams},
  {Hunten}, {Little}, {Saltzman}, {Alfred}, {O'Dougherty}, {Walthall},
  {Kenagy}, {Peterson}, {Crowther}, {Perry}, {See}, {Selznick}, {Sauve},
  {Beiser}, {Black}, {Pfisterer}, {Lancaster}, {Oliver}, {Oquest}, {Crowley},
  {Morgan}, {Castle}, {Dominguez}, and {Sullivan}}]{Rizk18}
{Rizk} B, {Drouet d'Aubigny} C, {Golish} D, {Fellows} C, {Merrill} C, {Smith}
  P, {Walker} MS, {Hendershot} JE, {Hancock} J, {Bailey} SH, {DellaGiustina}
  DN, {Lauretta} DS, {Tanner} R, {Williams} M, {Harshman} K, {Fitzgibbon} M,
  {Verts} W, {Chen} J, {Connors} T, {Hamara} D, {Dowd} A, {Lowman} A, {Dubin}
  M, {Burt} R, {Whiteley} M, {Watson} M, {McMahon} T, {Ward} M, {Booher} D,
  {Read} M, {Williams} B, {Hunten} M, {Little} E, {Saltzman} T, {Alfred} D,
  {O'Dougherty} S, {Walthall} M, {Kenagy} K, {Peterson} S, {Crowther} B,
  {Perry} ML, {See} C, {Selznick} S, {Sauve} C, {Beiser} M, {Black} W,
  {Pfisterer} RN, {Lancaster} A, {Oliver} S, {Oquest} C, {Crowley} D, {Morgan}
  C, {Castle} C, {Dominguez} R, {Sullivan} M (2018) {OCAMS: The OSIRIS-REx
  Camera Suite}. Space Science Reviews 214(1)

\bibitem[{{Russell} et~al.(2019){Russell}, {Miller-Jones}, {Sivakoff}, and
  {Tetarenko}}]{Russell19}
{Russell} TD, {Miller-Jones} JCA, {Sivakoff} GR, {Tetarenko} AJ (2019) {ATCA
  radio detection of the new X-ray transient MAXI J0637-430}. The Astronomer's
  Telegram 13275:1

\bibitem[{{Thayer} et~al.(2021){Thayer}, {Allen}, {Bautz}, {Binzel}, {Chodas},
  {Guevel}, {Hoak}, {Hong}, {Lambert}, {Masterson}, {Megerssa}, and
  {Ryu}}]{Thayer21}
{Thayer} C, {Allen} B, {Bautz} MW, {Binzel} RP, {Chodas} M, {Guevel} D, {Hoak}
  D, {Hong} J, {Lambert} M, {Masterson} R, {Megerssa} S, {Ryu} K (2021)
  {Performance of a Directly Deposited Optical Blocking Filter on X-ray CCDs:
  Case Study From the Regolith X-ray Imaging Spectrometer (REXIS) ExperimentS}.
  In Preparation

\bibitem[{{Tomsick} et~al.(2019){Tomsick}, {Garcia}, {Fabian}, {Walton},
  {Jiang}, {Fuerst}, {Buisson}, {Shaw}, {Hare}, {Bachetti}, {Connors},
  {Gandhi}, and {Xu}}]{Tomsick19}
{Tomsick} JA, {Garcia} J, {Fabian} A, {Walton} D, {Jiang} J, {Fuerst} F,
  {Buisson} D, {Shaw} A, {Hare} J, {Bachetti} M, {Connors} R, {Gandhi} P, {Xu}
  Y (2019) {A NuSTAR Observation of MAXI J0637-430: A New X-ray Transient and
  Likely Black Hole X-ray Binary}. The Astronomer's Telegram 13270:1

\bibitem[{{Trombka} et~al.(2000){Trombka}, {Squyres}, {Br{\"u}ckner},
  {Boynton}, {Reedy}, {McCoy}, {Gorenstein}, {Evans}, {Arnold}, {Starr},
  {Nittler}, {Murphy}, {Mikheeva}, {McNutt}, {McClanahan}, {McCartney},
  {Goldsten}, {Gold}, {Floyd}, {Clark}, {Burbine}, {Bhangoo}, {Bailey}, and
  {Petaev}}]{2000Sci...289.2101T}
{Trombka} JI, {Squyres} SW, {Br{\"u}ckner} J, {Boynton} WV, {Reedy} RC, {McCoy}
  TJ, {Gorenstein} P, {Evans} LG, {Arnold} JR, {Starr} RD, {Nittler} LR,
  {Murphy} ME, {Mikheeva} I, {McNutt} RL, {McClanahan} TP, {McCartney} E,
  {Goldsten} JO, {Gold} RE, {Floyd} SR, {Clark} PE, {Burbine} TH, {Bhangoo} JS,
  {Bailey} SH, {Petaev} M (2000) {The Elemental Composition of Asteroid 433
  Eros: Results of the NEAR-Shoemaker X-ray Spectrometer}. Science
  289(5487):2101--2105, \doi{10.1126/science.289.5487.2101}

\bibitem[{{Weisskopf} et~al.(2010){Weisskopf}, {Guainazzi}, {Jahoda},
  {Shaposhnikov}, {O'Dell}, {Zavlin}, {Wilson-Hodge}, and
  {Elsner}}]{Weisskopf10}
{Weisskopf} MC, {Guainazzi} M, {Jahoda} K, {Shaposhnikov} N, {O'Dell} SL,
  {Zavlin} VE, {Wilson-Hodge} C, {Elsner} RF (2010) {On Calibrations Using the
  Crab Nebula and Models of the Nebular X-Ray Emission}. The Astrophysical
  Journal 713(2):912--919, \doi{10.1088/0004-637X/713/2/912},
  \eprint{1003.1916}

\bibitem[{{Woods} et~al.(2017){Woods}, {Caspi}, {Chamberlin}, {Jones},
  {Kohnert}, {Mason}, {Moore}, {Palo}, {Rouleau}, {Solomon}, {Machol}, and
  {Viereck}}]{Woods17}
{Woods} TN, {Caspi} A, {Chamberlin} PC, {Jones} A, {Kohnert} R, {Mason} JP,
  {Moore} CS, {Palo} S, {Rouleau} C, {Solomon} SC, {Machol} J, {Viereck} R
  (2017) {New Solar Irradiance Measurements from the Miniature X-Ray Solar
  Spectrometer CubeSat}. The Astrophysical Journal Letters 835(2):122,
  \doi{10.3847/1538-4357/835/2/122}, \eprint{1610.01936}

\end{thebibliography}

%\begin{thebibliography}{}
% etc
%\end{thebibliography}

\appendix

\section{Angular Resolution and Point Source Localization in Coded-Aperture Imaging} \label{s:ang_resolution}

The angular resolution ($\Delta \theta$) 
and the localization error ($\delta \theta$) of a point source in coded-aperture imaging are given as 
\begin{eqnarray*}
    \Delta \theta &= & \tan^{-1}\left(\frac{\sqrt{\Delta m^2+ \Delta d^2}}{s}\right), \\
    \delta \theta & \sim & \frac{\Delta \theta}{\sigma},
\end{eqnarray*}
where $\Delta m$ is the mask pixel size, $\Delta d$ the detector pixel size, 
$s$ the separation from the mask and detector, and $\sigma$ the signal-to-noise ratio (SNR) of the point source. 

The upper limit in the localization error is set by the minimum SNR ($\gtrsim$5) needed to confidently claim
a detection. Coded aperture imagers with a random mask pattern such as REXIS also have a upper limit in the SNR they can achieve 
due to the {\it coding noise} arising from the random nature of the mask pattern. 

The upper limit in the SNR from a detector plane image is 
the square root of the number of mask pixels that the detector plane image can capture. 
For REXIS, the upper limit in SNR ranges from $\sim$ 18 to $\sim$ 25, 
depending the number of nodes used in the detector plane image.
This in turn sets the lower limit in the localization error. 
Even without the coding noise, there is a fundamental systematic  error in localization, 
which originates from the finite detector pixel size (25$''$) 
and the pointing jitter ($\sim10''$). 
Thus, the localization error of REXIS would be limited to $\gtrsim$ 30$''$, 
unless algorithms for the subpixel randomization and active jitter compensation are employed. 

\section{Modeling the CCD response of the Mn-\Ka line} \label{s:linemodel}

The response $f(E)$ of the 5.9 keV Mn-\Ka line in Figure~\ref{f:XIS_sim_spectra} can be
described by a Gaussian function and an exponential low-energy tail component:
\begin{eqnarray*}
    & f(E)  =  A_1  \exp\left(-\frac{(E - E_0)^2}{\Delta_1^2/\log16 }\right)  & \\
       & +  A_2  \bar{\theta}(E-E_0)  e^{C_0 (E - E_0)}  \left[ 1 - \exp\left(-\frac{(E - E_0)^2 }{ \Delta_2^2/ \log16} \right) \right] &, 
\end{eqnarray*}
where $A_1$ and $A_2$ are the normalization constants, $E_0$ the line energy, $\Delta_1$ and $\Delta_2$ the FWHM of the Gaussian and tail components,
$C_0$ the exponential decay parameter. $\bar{\theta}(x)$ is a reverse step function, where
\begin{eqnarray*}
    \bar{\theta}(x)  = \begin{cases}
    1,& \text{if } x \leq 0 \\
    0,              & \text{otherwise.}
\end{cases}
\end{eqnarray*}

\section{Pointing Coordinates of the Crab Nebula and Sco X-1 Observations} \label{s:pointings}

Tables~\ref{table:crab_observations} 
lists the boresight coordinates of the Crab Nebula observations during
the Crab Calibration Operation and the Sco X-1 observations during Mask Calibration Operation, respectively.
See Figure~\ref{f:crab_pointings}.

\begin{SCtable}
\small
\caption{Target boresight pointings during the Crab Nebula and Sco X-1 Observations. These
pointings were repeated on March 16, 17, 23, and 24 2019.}
\begin{tabular}{ cccc }
\hline\hline
Order & R.A. & Dec & Offset\\
\hline
\multicolumn{4}{l}{The Crab Nebula: 2019 March 16, 17, 23, and 24}\\
1 & 83$^{\circ}$ 37' 48"  & 22$^{\circ}$ 00' 36"  & 0.0$^\circ$\\
2 & 80$^{\circ}$ 43' 12"  & 19$^{\circ}$ 48' 00"  & 3.5$^\circ$\\
3 & 83$^{\circ}$ 37' 48"  & 18$^{\circ}$ 30' 36"  & 3.5$^\circ$\\
4 & 86$^{\circ}$ 32' 24"  & 19$^{\circ}$ 48' 00"  & 3.5$^\circ$\\
5 & 87$^{\circ}$ 20' 24"  & 22$^{\circ}$ 44' 24"  & 3.5$^\circ$\\
6 & 85$^{\circ}$ 19' 12"  & 25$^{\circ}$ 09' 00"  & 3.5$^\circ$\\
7 & 81$^{\circ}$ 58' 12"  & 25$^{\circ}$ 09' 36"  & 3.5$^\circ$\\
8 & 79$^{\circ}$ 55' 48"  & 22$^{\circ}$ 45' 36"  & 3.5$^\circ$\\
\hline
\multicolumn{4}{l}{Sco X-1: 2019 June 2 }\\
1 & 244$^{\circ}$ 58' 44"  & -15$^{\circ}$ 38' 24"  & 0.0$^\circ$\\
2 & 232$^{\circ}$ 00' 00"  & -15$^{\circ}$ 38' 24"  & 12.5$^\circ$\\
3 & 250$^{\circ}$ 00' 00"  & -15$^{\circ}$ 38' 24"  & 4.8$^\circ$\\
4 & 244$^{\circ}$ 58' 44"  & -21$^{\circ}$ 38' 24"  & 6.0$^\circ$\\
5 & 240$^{\circ}$ 00' 00"  & -04$^{\circ}$ 38' 24"  & 12.0$^\circ$\\
\hline
\end{tabular}
\label{table:crab_observations}
\end{SCtable}

\end{document}